\documentclass[preprint2]{aastex631}

\graphicspath{{./}{figures/}}

\usepackage[all]{hypcap}

\hypersetup{
    colorlinks=true,       
    linkcolor=blue,                 
    filecolor=blue,      
    urlcolor=blue
}

\usepackage{lineno}
\begin{document}

\title{Gamma rays from Fast Black-Hole Winds}

%author list:
\author{M.~Ajello}
\email{majello@clemson.edu}
\affiliation{Department of Physics and Astronomy, Clemson University, Kinard Lab of Physics, Clemson, SC 29634-0978, USA}
\author{L.~Baldini}
\affiliation{Universit\`a di Pisa and Istituto Nazionale di Fisica Nucleare, Sezione di Pisa I-56127 Pisa, Italy}
\author{J.~Ballet}
\affiliation{AIM, CEA, CNRS, Universit\'e Paris-Saclay, Universit\'e de Paris, F-91191 Gif-sur-Yvette, France}
\author{G.~Barbiellini}
\affiliation{Istituto Nazionale di Fisica Nucleare, Sezione di Trieste, I-34127 Trieste, Italy}
\affiliation{Dipartimento di Fisica, Universit\`a di Trieste, I-34127 Trieste, Italy}
\author{D.~Bastieri}
\affiliation{Istituto Nazionale di Fisica Nucleare, Sezione di Padova, I-35131 Padova, Italy}
\affiliation{Dipartimento di Fisica e Astronomia ``G. Galilei'', Universit\`a di Padova, I-35131 Padova, Italy}
\author{R.~Bellazzini}
\affiliation{Istituto Nazionale di Fisica Nucleare, Sezione di Pisa, I-56127 Pisa, Italy}
\author{A.~Berretta}
\affiliation{Dipartimento di Fisica, Universit\`a degli Studi di Perugia, I-06123 Perugia, Italy}
\author{E.~Bissaldi}
\affiliation{Dipartimento di Fisica ``M. Merlin" dell'Universit\`a e del Politecnico di Bari, via Amendola 173, I-70126 Bari, Italy}
\affiliation{Istituto Nazionale di Fisica Nucleare, Sezione di Bari, I-70126 Bari, Italy}
\author{R.~D.~Blandford}
\affiliation{W. W. Hansen Experimental Physics Laboratory, Kavli Institute for Particle Astrophysics and Cosmology, Department of Physics and SLAC National Accelerator Laboratory, Stanford University, Stanford, CA 94305, USA}
\author{E.~D.~Bloom}
\affiliation{W. W. Hansen Experimental Physics Laboratory, Kavli Institute for Particle Astrophysics and Cosmology, Department of Physics and SLAC National Accelerator Laboratory, Stanford University, Stanford, CA 94305, USA}
% ...

\begin{abstract}

Massive black holes at the centers of galaxies can launch powerful wide-angle winds that, if sustained over time, can unbind the gas from the stellar bulges of galaxies. These winds may be responsible for the observed scaling relation between the masses of the central black holes and the velocity dispersion of stars in galactic bulges. Propagating through the galaxy, the wind should interact with the interstellar medium creating a strong shock, similar to those observed in supernovae explosions, which is able to accelerate charged particles to high energies. In this work we use data from the {\it Fermi} Large Area Telescope to search for the $\gamma$-ray emission from galaxies with an ultra-fast outflow (UFO): a fast ($v\sim0.1$c), highly ionized outflow, detected in absorption at hard X-rays in several nearby active galactic nuclei (AGN). Adopting a sensitive stacking analysis we are able to detect the average $\gamma$-ray emission from these galaxies and exclude that it is due to processes other than the UFOs. Moreover, our analysis shows that the $\gamma$-ray luminosity scales  with the AGN bolometric luminosity and that these outflows transfer $\sim$0.04\,\% of their mechanical power to $\gamma$ rays. Interpreting the observed $\gamma$-ray emission as produced by cosmic rays (CRs) accelerated at the shock front, we find that the $\gamma$-ray emission may attest to the onset of the wind-host interaction and that these outflows can energize charged particles up to the transition region between galactic and extragalactic CRs.

\end{abstract}

\section{Introduction} \label{sec:intro}

Accreting super-massive black holes (SMBHs) at the centers of galaxies, often called  active galactic nuclei (AGN), have been observed to launch and power outflows, which can have a dramatic impact on the host galaxies themselves, the intergalactic medium, and the intracluster medium~\citep{Silk:1997xw,McNamara:2007ww,Somerville:2008bx,McCarthy:2009kk,Hopkins:2009yg}. One spectacular, well observed, type of outflow are relativistic jets, where particles are accelerated to near the speed of light in narrow collimated beams (often with an opening angle of $\sim1^\circ$), which can extend up to Mpc scales. These relativistic jets shine at all wavelengths, but are easily studied in radio, X-rays, and $\gamma$ rays when the jet axis is not far from our line of sight. Black-hole winds~\citep{king2015}, on the other hand, are AGN outflows that are not collimated and are generally more difficult to detect, although no less important. Indeed, AGN winds have been proposed as the mechanism able to regulate the co-evolution of the galaxy and its central SMBH, which is observed in the scaling of the black-hole mass and the bulge velocity dispersion~\citep{gebhart2000,ferrarese2005,kormendy2013}. AGN winds that are powerful enough can heat up and eject the gas from the galaxy, regulating the growth of both the galaxy itself and the black hole.

The most powerful AGN winds can reach velocities of $\sim$0.1$-$0.3c~\citep{chartas2002,pounds2003,reeves2003massive,tombesi2010discovery} and can carry enough energy to unbind the gas of the stellar bulge~\citep{king2015}. Some of these winds have been identified in nearby AGN through X-ray observations of blue-shifted Fe K-shell absorption lines~\citep{reeves2003massive,tombesi2010discovery,tombesi2010evidence,tombesi2012evidence,gofford2013suzaku}.

These winds, which have been dubbed ultra-fast outflows (UFOs), are made of highly ionized gas and are likely launched from near the SMBH~\citep{king2003}. Their wide solid angle [$\Omega/2\pi\approx0.4$, \citep{gofford2015suzaku}] and fast velocity allow UFOs to transfer a significant amount of kinetic energy from the AGN to the host galaxy. They are also believed to be common in nearby AGN~\citep{king2015}.

UFOs, while traveling outward, interact and shock the interstellar medium \citep[ISM,][]{king2010}, producing a reverse shock and a forward shock. The reverse shock decelerates the wind itself while the forward shock travels through the galaxy with a velocity in the $\sim$200-1000\,km s$^{-1}$ range and leads to the formation of a bubble of hot, tenuous gas, e.g.,~\citet{zubovas2012}. Because of the cooling, the phase and velocity of the outflow should change, eventually leading to the formation of low-velocity molecular outflows, commonly observed in many ultra-luminous infrared galaxies \citep[see e.g.][]{cicone2014,feruglio2015}.
Indeed, there are a handful of objects like IRAS 17020+4544~\citep{longinotti2018early} and Mrk 231~\citep{feruglio2015} where both a UFO and molecular outflow have been detected and found in agreement with the prediction of the energy-conserving outflow model, which is the basis of AGN feedback~\citep{fabian2012}.

UFOs have velocities comparable to (or even larger than) those of the ejecta launched in  supernova explosions, which are known to shock the ISM and accelerate cosmic rays (CRs).
Gamma-ray emission is a signature of the interaction of relativistic charged particles with ambient gas and photon fields and has been observed in many cases in supernova remnants~\citep{Acero:2015prw}. Given the similarity, in this work we search for the $\gamma$-ray emission from UFOs using the Large Area Telescope \citep[LAT][]{atwood2009}] on board the {\it Fermi Gamma-ray Space Telescope}~\citep{Atwood:2009ez}.

Models of the $\gamma$-ray emission from AGN outflows \citep{wang2016contribution,lamastra2017extragalactic}
show them to be weak emitters, with $\gamma$-ray luminosities of $\approx 10^{40} \ \mathrm{erg \ s^{-1}}$, which explains why UFOs have not yet been detected by the LAT\footnote{No $\gamma$-ray source from the 4FGL catalog~\citep{4FGL} is a associated to a UFO.}. Here, we adopt a different strategy and search for the collective $\gamma$-ray emission from a sample of UFOs using a stacking technique. 

The paper is organized as follows. In $\S$~\ref{sec:sample_selection} and $\S$~\ref{sec:data_and_model} we describe the sample selection and the data analysis. Results are presented in $\S$~\ref{sec:results}, with additional tests discussed in $\S$~\ref{sec:additional_tests}.  $\S$~\ref{sec:sed_modeling} reports the theoretical interpretation for the  observed  $\gamma$-ray emission, while a discussion is reported in $\S$~\ref{sec:discussion}. Finally, $\S$~\ref{sec:summary_and_conclusion} reports our conclusions.

\section{Sample Selection}
\label{sec:sample_selection}
We start from  a sample of 35 sources {that} have been identified as UFOs through X-ray observations~\citep{reeves2003massive,tombesi2010discovery,tombesi2010evidence,tombesi2012evidence,gofford2013suzaku}. We have verified that none of the objects are positionally coincident with any known $\gamma$-ray  sources {reported in the 4FGL \citep{4FGL}}. From the initial sample we make the following cuts. First, we only keep the radio-quiet sources (as specified in the original references) to avoid contamination of the signal from the relativistic jet. Furthermore, we only select sources that are nearby ($z<0.1$) with a mildly relativistic wind velocity ($v>0.1c$). The former cut is motivated by the expected low luminosity of the UFO emission~\citep{wang2016contribution}, and the latter cut is motivated by the fact that the $\gamma$-ray emission is predicted to scale with the kinetic power of the outflow~\citep{wang2016contribution,lamastra2017extragalactic}. After making these cuts we are left with 11 sources, which we use as our benchmark sample. The details of these sources are reported in Table~\ref{tab:ufo_parameters}.

\begin{figure}[t]

\includegraphics[width=0.48\textwidth]{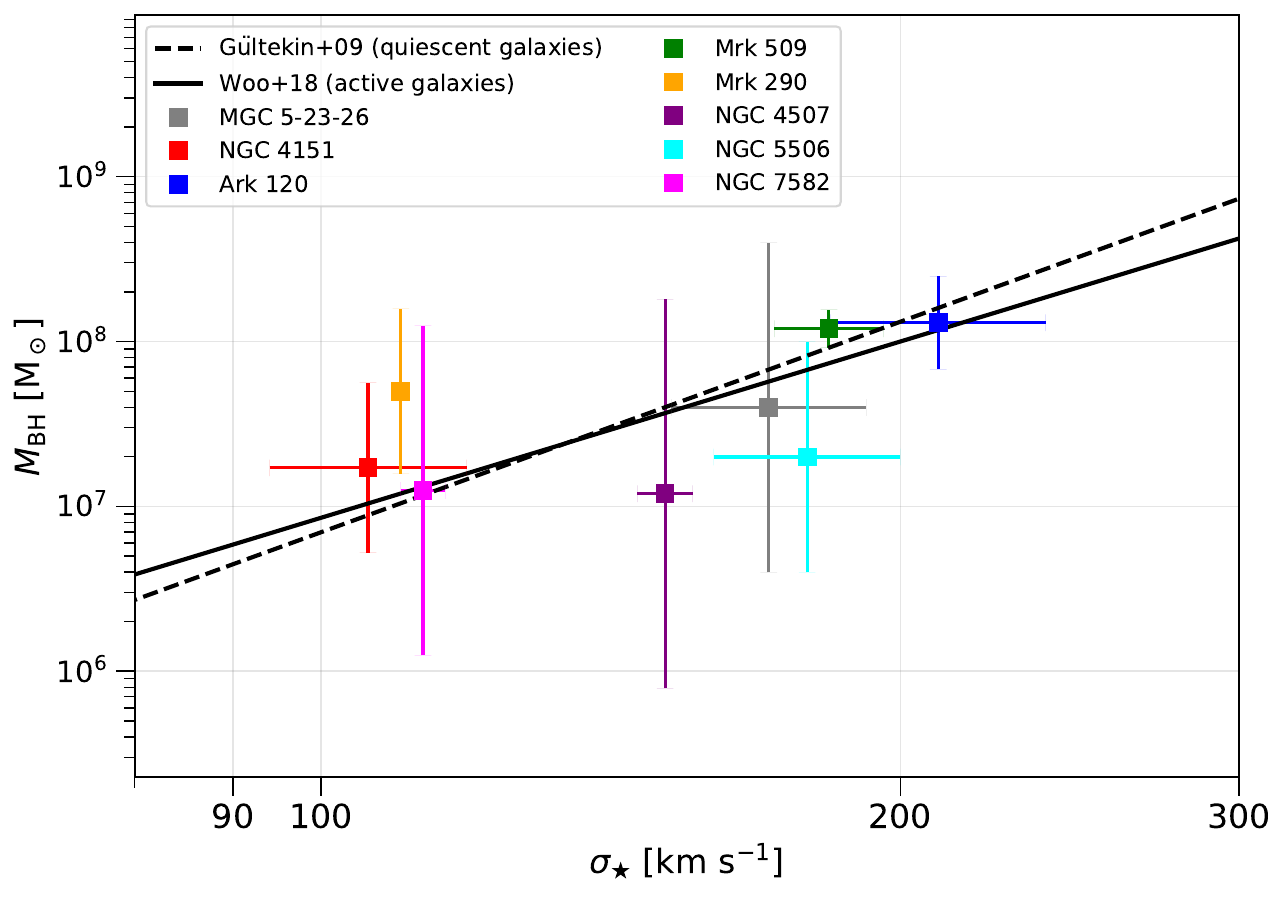} 
\caption{Bulge stellar velocity dispersion versus black-hole mass for our UFO sample, with values taken from the literature. Measurements were found for 8/11 sources. The error bars are statistical plus systematic, where the systematic uncertainty comes from different independent estimates. Information for the velocity dispersion measurements is provided in Table~\ref{tab:velocity_dispersion}. To quantify the systematic uncertainty in the black-hole mass, we use minimum and maximum values from the different references provided in Table~\ref{tab:velocity_dispersion}, as well as the values given in Table~\ref{tab:ufo_parameters}. The solid and dashed lines show the scaling relations for active and quiescent galaxies, from~\cite{2010ApJ...716..269W} and~\cite{2009ApJ...698..198G}, respectively.}
\label{fig:mass_disp}

\end{figure}

Table~\ref{tab:velocity_dispersion} reports additional properties of our sample of UFOs, including the bulge velocity dispersion,  1.4\,GHz radio flux and total (8-1000\,$\mu$m) IR luminosity. Figure~\ref{fig:mass_disp} shows that the UFOs considered here obey the M-$\sigma$ relation well~\citep{2009ApJ...698..198G,2010ApJ...716..269W}, strengthening the  evidence that these outflows operated in the energy-conserving phase in the past~\citep{king2015}. Finally, the origin of the radio emission in radio-quiet AGN is not very clear and it is likely due to a number of phenomena including AGN winds, star formation, free-free emission from photo-ionized gas and AGN coronal activity~\citep{panessa2019}. For these reasons, the radio fluxes reported in Table~\ref{tab:velocity_dispersion} are interpreted as upper limits to the synchrotron emission from accelerated electrons, as discussed in Section~\ref{sec:sed_modeling}. 

We note that there are alternative models explaining the absorption features as produced not by an outflowing wind, but as resonant absorption by highly ionized iron in the accretion disk~\citep{gallo2011}. However,  this model has difficulties explaining several of the observed properties of the UFO features like  the presence of P-Cygni profiles~\citep{nardini2015,chartas2016}, or the correlation between outflow velocity and the AGN bolometric luminosity~\citep{saez2011,matzeu2017}.

\begin{deluxetable*}{lcccccccccc}
\tabletypesize{\scriptsize}
\tablecaption{UFO Source Sample}\label{tab:ufo_parameters}
\tablewidth{0pt}
\tablehead{
\colhead{Name} & \colhead{RA $(^{\circ})$} & \colhead{DEC $(^{\circ})$} & \colhead{Type} & \colhead{Redshift} & \colhead{Velocity} & \colhead{log$M_{\mathrm{BH}}$} & \colhead{log$\dot{E}^{\mathrm{Min}}_K$} & \colhead{log$\dot{E}^{\mathrm{Max}}_K$} & \colhead{log$L_{\mathrm{Bol}}$} & \colhead{95\% UL ($\times 10^{-11}$) } \\
\colhead{} & \colhead{$[{\rm J2000}]$} & \colhead{$[{\rm J2000}]$} & \colhead{} & \colhead{$[z]$} & \colhead{$[v/c]$} & \colhead{$[M_\odot]$} & \colhead{$[\mathrm{erg \ \mathrm{s}^{-1}}]$} & \colhead{$[\mathrm{erg \ \mathrm{s}^{-1}}]$} & \colhead{$[\mathrm{erg \ \mathrm{s}^{-1}}]$} & \colhead{$\mathrm{[ph \ cm^{-2} \ s^{-1}]}$}}
\decimalcolnumbers
\startdata
Ark 120$^{\emph{a,c}}$ & 79.05 & $-$0.15 & Sy1 & 0.033 & 0.27  &8.2 $\pm$ 0.1 &$>43.1$ & 46.2 $\pm$ 1.3 &45.0$^{\emph{f}}$ & 7.5 \\
&&&&&&&&&44.2$^{\emph{h}}$&\\
&&&&&&&&&\textbf{44.6}&\\
MCG-5-23-16$^{\emph{a,c}}$ & 146.92 & $-$30.95 & Sy2 & 0.0084 & 0.12  &7.6 $\pm$ 1.0 & 42.7 $\pm$ 1.0 & 44.3 $\pm$ 0.2 &44.1$^{\emph{k}}$ & 4.3 \\
NGC 4151$^{\emph{a,c}}$ & 182.64 & 39.41 & Sy1 &0.0033 & 0.105 &7.1 $\pm$ 0.2 &$>$41.9 & 43.1 $\pm$ 0.5 &44.1$^{\emph{g}}$ & 10.6  \\
&&&&&&&&&42.9$^{\emph{h}}$&\\
&&&&&&&&&43.9$^{\emph{i}}$&\\
&&&&&&&&&42.9$^{\emph{j}}$&\\
&&&&&&&&&43.2$^{\emph{j*}}$&\\
&&&&&&&&&\textbf{43.4}&\\
PG 1211+143$^{\emph{a,c}}$  & 183.57 & 14.05 & Sy1 & 0.081 & 0.13 &8.2 $\pm$ 0.2 & 43.7 $\pm$ 0.2 & 46.9 $\pm$ 0.1 &45.7$^{\emph{f}}$ & 3.7 \\
&&&&&&&&&44.8$^{\emph{h}}$&\\
&&&&&&&&&44.7$^{\emph{j}}$&\\
&&&&&&&&&45.0$^{\emph{j*}}$&\\
&&&&&&&&&\textbf{45.1}&\\
NGC 4507$^{\emph{a,c}}$ & 188.90 & $-$39.91 & Sy2 & 0.012 & 0.18 &6.4 $\pm$ 0.5 &$>41.2$ & 44.6 $\pm$ 1.1 &44.3$^{\emph{e}}$ & 3.4 \\
NGC 5506$^{\emph{b,d}}$ & 213.31 & $-$3.21 & Sy1.9 & 0.006 & 0.25 &7.3 $\pm$ 0.7  & 43.3 $\pm$ 0.1 & 44.7 $\pm$ 0.5 & 44.3$^{\emph{e}}$ & 6.4 \\
Mrk 290$^{\emph{a,c}}$ & 233.97 & 57.90 &Sy1 & 0.030 & 0.14 &7.7 $\pm$ 0.5 &43.4 $\pm$ 0.9 & 45.3 $\pm$ 1.2 &44.4$^{\emph{e}}$ & 4.5 \\
Mrk 509$^{\emph{a,c}}$ & 311.04 & $-$10.72 & Sy1 & 0.034 & 0.17 &8.1 $\pm$ 0.1 &$>$43.2 & 45.2 $\pm$ 1.0 &45.2$^{\emph{e}}$ & 9.5 \\
&&&&&&&&&44.3$^{\emph{h}}$&\\
&&&&&&&&&45.3$^{\emph{i}}$&\\
&&&&&&&&&44.3$^{\emph{j}}$&\\
&&&&&&&&&44.5$^{\emph{j*}}$&\\
&&&&&&&&&\textbf{44.7}&\\
SWIFT~J2127.4+5654$^{\emph{b,d}}$ & 321.94 & 56.94 & Sy1 & 0.014 & 0.23 & $\sim$7.2 &42.8 $\pm$ 0.1 & 45.6 $\pm$ 0.5 &44.5$^{\emph{d}}$ & 9.1 \\
MR 2251-178$^{\emph{b,d}}$ & 343.52 & $-$17.58& Sy1 & 0.064 & 0.14 &8.7 $\pm$ 0.1 &43.3 $\pm$ 0.1 & 46.7 $\pm$ 0.7 &45.8$^{\emph{f}}$ & 7.4 \\
NGC 7582$^{\emph{a,c}}$ & 349.60 & $-$42.37 &Sy2 & 0.0052 & 0.26 &7.1 $\pm$ 1.0 &43.4 $\pm$ 1.1 & 44.9 $\pm$ 0.4 &43.3$^{\emph{e}}$ & 4.7 \\
\enddata
\tablecomments{Our sample is comprised of 11 sources with $z<0.1$ and $v>0.1c$. The first superscript on the source name indicates the reference for the detection, and the second superscript indicates the reference for the UFO parameters (columns $6-9$), where $\dot{E}^{\mathrm{Min}}_K$ and $\dot{E}^{\mathrm{Max}}_K$ are the minimum and maximum kinetic powers.   Values for the bolometric luminosity $(L_{\mathrm{Bol}})$ are taken from the literature, with the reference indicated by the superscript. For sources with numerous determinations we also give the mean value in boldface text. The $\gamma$-ray flux ($1-800$ GeV) upper limit (UL) is calculated at the 95\% confidence level, using a photon index of --2.0. 
	    $^{\emph{a}}$~\citet{tombesi2010evidence}; $^{\emph{b}}$~\citet{gofford2013suzaku}; $^{\emph{c}}$~\citet{tombesi2012evidence}; $^{\emph{d}}$~\citet{gofford2015suzaku}; $^{\emph{e}}$~\citet{vasudevan2010power}; $^{\emph{f}}$~\citet{Vasudevan:2007hz}; $^{\emph{g}}$~\citet{Vasudevan:2008wn}; $^{\emph{h}}$~\citet{Peterson:2004nu}; $^{\emph{i}}$~\citet{crenshaw2012feedback}; $^{\emph{j}}$~\citet[5100 $\AA$ flux density]{Kaspi:2005wx}; $^{\emph{j*}}$~\citet[1450 $\AA$ flux density]{Kaspi:2005wx}; $^{\emph{k}}$~\citet{alonso2011torus}.}
\end{deluxetable*}

%%%%%%%%%%%%%%%%%%%%%%%%%%%%%%%%%%%%%%%%%%%%%
%%%%%%%%%%%%%%%%%%%%%%%%%%%%%%%%%%%%%%%%%%%%%%

\begin{deluxetable*}{lccc}
\tabletypesize{\scriptsize}
\tablecaption{Additional UFO Properties}\label{tab:velocity_dispersion}
\tablewidth{0pt}
\tablehead{
\colhead{Name} & \colhead{Velocity Dispersion} & \colhead{1.4 GHz Radio Flux$^{\emph{g}}$} & \colhead{IR Lum.$^{\emph{h}}$} \\
\colhead{} & \colhead{$[\mathrm{km/s}]$} & \colhead{[mJy]} & \colhead{log (L$_{\odot}$)}
}
\decimalcolnumbers
\startdata
Ark 120 & 184, 238$^{\emph{a,b}}$ & 12.4 & 11.0\\
MCG-5-23-16 & 152, 192$^{\emph{a,c}}$ & 14.3  & 9.6\\
NGC 4151 & 94, 119$^{\emph{a,c}}$  &  347.6 & 10.2\\
PG 1211+143 & \nodata & 4.3 & \nodata\\
NGC 4507$^{\emph{*}}$ & 146, 156$^{\emph{d}}$  & 67.4 & 10.5\\
NGC 5506 & 160, 200$^{\emph{d}}$  & 355 & 10.5 \\
Mrk 290 & 109, 111$^{\emph{e}}$  & 5.32 & $<$10.3\\
Mrk 509 & 172, 196$^{\emph{b}}$  & 19.2 & 10.5\\
SWIFT~J2127.4+5654 & \nodata  & 6.4 & 10.4\\
MR 2251-178 & \nodata  & 16 & $<10.5$\\
NGC 7582 & 110, 116$^{\emph{d}}$ &270 & 10.6  \\
\enddata
\tablecomments{The second column gives velocity dispersion measurements taken from the literature, with the references indicated by the superscripts. Measurements were found for 8/11 sources, and we provide minimum and maximum values (separated by a comma). For sources with just one reference, the range is due to statistical error only, and for sources with two references, the range also includes the systematic error due to the different estimates.
$^{\emph{*}}$~Note that most published estimates of the black-hole mass for NGC 4507 are based on velocity dispersion and [O III] line widths, and thus they are not independent measures. In quantifying the uncertainty in Figure~\ref{fig:mass_disp} we also use black-hole mass values from~\citet{bian2007eddington,beifiori2012correlations,nicastro2003lack}.
$^{\emph{a}}$~\citet{2010ApJ...716..269W};  
$^{\emph{b}}$~\citet{Grier:2013sza};
$^{\emph{c}}$~\citet{2014ApJ...791...37O}; 
$^{\emph{d}}$~\citet{2012ApJ...748..130M};
$^{\emph{e}}$~\citet{2015ApJ...809...20B};
$^{\emph{f}}$~Hyperleda; 
$^{\emph{g}}$~NVSS~\citep{condon_1998};
$^{\emph{h}}$~IRAS~\citep{kleinmann86,moshir90}.}
\end{deluxetable*}

\section{Data Analysis}
\label{sec:data_and_model}
\subsection{Data}

We analyze  data collected by \textit{Fermi}-LAT between 2008 August 04 to 2019 September 10 (11.1 years). The events have energies in the range 1$-$800 GeV and are binned in 8 bins per decade. The pixel size is $0.08^\circ$. To reduce contamination from the Earth's limb we use a maximum zenith angle of $105^\circ$. {We define a $10^\circ \times 10^\circ$ region of interest (ROI) centered at the position of each UFO source.} We use the standard data filters: DATA\_QUAL$>$0 and LAT\_CONFIG==1. The analysis is performed using Fermipy (v0.18.0)\footnote{Available at \url{https://fermipy.readthedocs.io/en/latest/}}, which utilizes the underlying Fermitools (v1.2.23). 

We select photons corresponding to the P8R3\_SOURCE\_V2 class~\citep{Atwood2013}. In order to optimize the sensitivity of our stacking technique we implement a joint likelihood analysis with the four point spread function (PSF) event types available in the Pass 8 data set\footnote{For more information on the different PSF types see \url{https://fermi.gsfc.nasa.gov/ssc/data/analysis/documentation/Cicerone/Cicerone_Data/LAT_DP.html}.}. The data is divided into quartiles corresponding to the quality of the reconstructed direction, from the lowest quality quartile (PSF0) to the best quality quartile (PSF3). Each sub-selection has its own binned likelihood instance that is combined in a global likelihood function for the ROI. This is easily implemented in Fermipy by specifying the components section in the configuration file. Each PSF type also has its own corresponding isotropic spectrum, namely, iso\_P8R3\_SOURCE\_V2\_PSF{\it i}\_v1, for $i$ ranging from 0$-$3. The Galactic diffuse emission is modeled using the standard component (gll\_iem\_v07), and the point source emission is modeled using the 4FGL catalog (gll\_psc\_v20). In order to account for photon leakage from sources outside of the ROI due to the PSF of the detector, the model includes all 4FGL sources within a $15^\circ \times 15^\circ$ region. The energy dispersion correction (edisp\_bins=--1) is enabled for all sources except the isotropic component.

\subsection{Analysis}
In the local Universe ($z<0.1$) UFOs are predicted to have a $\gamma$-ray luminosity of $\sim 10^{40} \ \mathrm{erg \ s^{-1}}$~\citep{wang2016contribution}, making them too faint to be detected individually by \textit{Fermi}-LAT. {Indeed, adopting the average photon index in the 4FGL of $\Gamma=-2.2$ we derive a $>1$\, GeV flux of $3.3\times10^{-12}$\, ph cm$^{-2}$ s$^{-1}$, for a source with a luminosity of 10$^{40}$\,erg s$^{-1}$ at $z=0.014$ (the median redshift of our sample). This flux is  $\sim$2.5 times fainter than the weakest source reported in the 4FGL.} We therefore analyze our source sample using a stacking technique. {This} technique has been developed previously  and has been successfully employed for multiple studies, i.e. upper limits on dark matter interactions~\citep{lat_2011_dwarfs}, detection of the extragalactic background light~\citep{Ajello:2018sxm}, extreme blazars~\citep{paliya2019fermi}, and star-forming galaxies~\citep{Ajello:2020zna}. 

The main assumption that we make for the stacking technique is that the sample of UFOs we are considering can be characterized by average quantities like the average flux and the average photon index (when we model their spectra with a power law). There are then two steps to the method. In the first step the model components are optimized for each ROI using a maximum likelihood fit. We evaluate the significance of each source in the ROI using the TS, which is defined as:
 
\begin{equation}\label{eq:TSeq}
     \mathrm{TS} = -2\mathrm{log}(\mathrm{L_0/L}),
\end{equation}

 \noindent where $\mathrm{L_0}$ is the likelihood for the null hypothesis, and L is the likelihood for the alternative hypothesis\footnote{For a more complete explanation of the TS resulting from a likelihood fit see~\cite{1996ApJ...461..396M} and \url{https://fermi.gsfc.nasa.gov/ssc/data/analysis/documentation/Cicerone/Cicerone\_Likelihood/}.}.
 For the first iteration of the fit, the spectral parameters of the Galactic diffuse component (index and normalization) and the isotropic component are freed. In addition, we free the normalizations of all 4FGL sources with TS$\geq$25 that are within $5^\circ$ of the ROI center, as well as sources with TS$\geq$500 and within $7^\circ$. Lastly, the UFO source is fit with a power-law spectral model, and the spectral parameters (normalization and index) are also freed. In the first step we also find new point sources using the Fermipy function \textit{find\_sources}, which generates TS maps and identifies new sources based on peaks in the TS. The TS maps are generated using a power-law spectral model with an index of $-2.0$. The minimum separation between two point sources is set to $0.5^\circ$, and the minimum TS for including a source in the model is set to 16.   

In the second step 2D TS profiles are generated for the spectral parameters of each UFO source, where the TS is defined as in Eq.~\ref{eq:TSeq}. We scan photon indices from --1 to --3.3 with a spacing of 0.1 and total integrated photon flux (between 1--800 GeV) from $10^{-13}$ to $10^{-9}$ $\mathrm{ph \ cm^{-2}\ s^{-1}}$ with 40 logarithmically spaced bins, freeing just the parameters of the diffuse components. For this step the power-law spectra of the UFOs are defined in terms of the total flux ($F_{\rm tot}$), integrated between the minimum energy ($E_{\mathrm{min}}$) and the maximum energy ($E_{\mathrm{max}}$):

\begin{equation}\label{eq:PLmodel}
\frac{dN}{dE} = \frac{F_{\rm tot}(\Gamma + 1)E^\Gamma}{E^{\Gamma+1}_{\mathrm{max}} - E^{\Gamma+1}_{\mathrm{min}}}
\end{equation}
Note that the likelihood value for the null hypothesis is calculated at the end of the first step by removing the UFO source from the model. Since we perform a joint likelihood in the different PSF event types (PSF0 $-$ PSF3), the total profile for each source is obtained by adding the profiles from each  of the four event types. Lastly, the TS profiles for all sources are added to obtain the stacked profile.  The TS is an additive quantity, and so the stacked profile gives the statistical significance for the combined signal.  

We validated the stacking method relying on a set of Monte Carlo simulations that reproduce the \textit{Fermi}-LAT observations. In these tests, the simulations include the isotropic and Galactic emission, as well as an isotropic population of point sources resembling blazars, which account for the vast majority of sources detected by \textit{Fermi}-LAT. Faint, below-threshold ``blazars" are included in the synthetic sky following the models of~\citet{Ajello:2015mfa}. Using this setup two different tests were performed. The stacking analysis was performed at 60 random ``empty" positions, i.e., positions away from bright detected sources. This analysis yielded no detection, confirming that the technique does not generate spurious detections. The second set of tests was aimed at characterizing the detected signal. The stacking was performed for 60 simulated sources whose flux was extracted from a power-law distribution with index $-2.5$ and minimum and median flux of respectively $4 \times 10^{-10} \ \mathrm{ph \ cm^{-2} \  s^{-1}}$ and $6.4 \times 10^{-10} \ \mathrm{ph \ cm^{-2} \  s^{-1}}$. The photon indices were extracted from a Gaussian distribution with average 
$-$2.21 and dispersion of 0.2. The values derived from the stacking analysis (flux =$7.0^{+0.6}_{-0.7} \times 10^{-10} \  \mathrm{ph \ cm^{-2} \  s^{-1}}$ and index of $-2.24 \pm 0.05$) are in agreement with the inputs, showing that our analysis successfully retrieves the average quantities of a population of sources. Moreover, the likelihood profile would not show a significant peak if those average quantities were not representative of the population.

\section{Results}
\label{sec:results}
\subsection{Stacked TS Profile for The Benchmark Sample}
The log-likelihoods (i.e.~logL) are maximized with the optimizer \textit{MINUIT}~\citep{james1975minuit}, and we have verified that each fit converges properly, as indicated by the \textit{MINUIT} outputs of $\mathrm{quality} = 3$ and $\mathrm{status} = 0$. The 95\% flux upper limits from the preprocessing step are reported in Table~\ref{tab:ufo_parameters}. 

\begin{figure}[t]
\begin{center}
\includegraphics[width=0.48\textwidth]{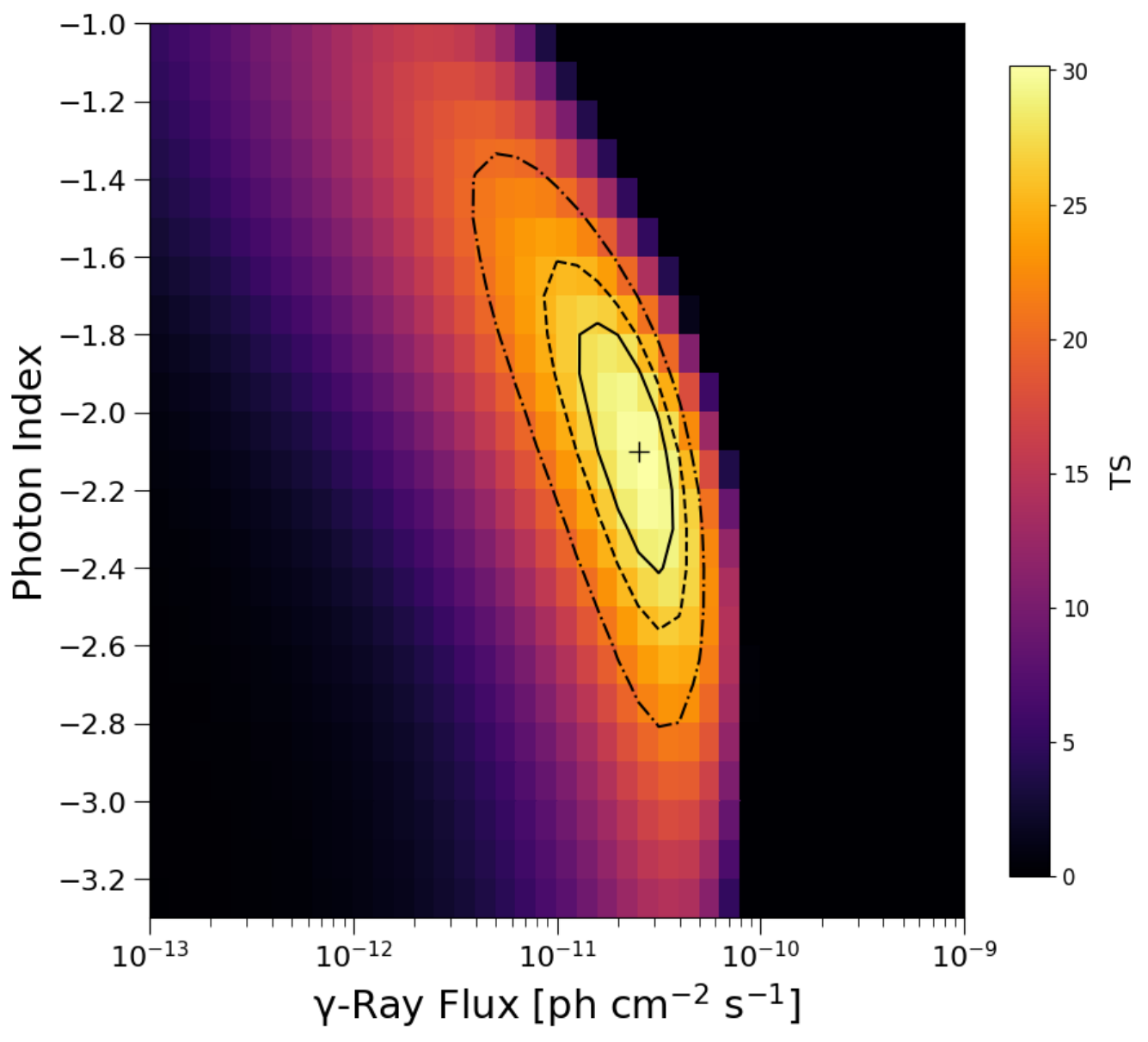} 
\caption{Stacked TS profile for the sample of UFOs. The color scale indicates the TS, and the plus sign indicates the location of the maximum value, with a $\mathrm{TS} = 30.1$ ($5.1\sigma$). Significance contours (for 2 degrees of freedom) are overlaid on the plot showing the 68\%, 90\%, and 99\% confidence levels, corresponding to $\Delta \ \mathrm{TS}$ = 2.30, 4.61, and 9.21, respectively.}
\label{fig:Run_5_JL}
\end{center}
\end{figure}

The stacked profile for our UFO sample is shown in Figure~\ref{fig:Run_5_JL}. The maximum TS is 30.1
(5.1$\sigma$)\footnote{The conversion from TS to $\sigma$ has been performed on the assumption that the TS behaves asymptotically as a $\chi^2$ distribution with 2 d.o.f~\citep{Mattox:1996zz}. Additionally, the Akaike information criterion test also shows the null hypothesis to be highly disfavored with a relative likelihood of $2\times10^{-6}$.}
, corresponding to a best-fit index of $-2.1\pm 0.3$ and a best-fit photon flux (1$-$800 GeV) of $2.5^{+1.5}_{-0.9} \times 10^{-11} \ \mathrm{ph \ cm^{-2} \ {s^{-1}}}$. 
The 68\%, 90\%, and 99\% significance contours are overlaid on the map, and as can be seen the spectral parameters are well constrained. The source with the overall highest individual TS is NGC 4151, having a maximum value of 21.2 (4.2\,$\sigma$), corresponding to a best-fit index of $-1.9^{+0.5}_{-0.3}$ and a best-fit flux of $6.3^{+3.7}_{-3.8} \times 10^{-11} \ \mathrm{ph \ cm^{-2} \ {s^{-1}}}$. The stacking analysis excluding NGC~4151 yields a maximum TS of 15.1 ($3.5\,\sigma$), corresponding to a best-fit index of $-2.2 \pm 0.4$ and a best-fit flux of $2.0^{+2.0}_{-1.0} \times 10^{-11} \ \mathrm{ph \ cm^{-2} \ {s^{-1}}}$.  

%To verify that the combined signal is not dominated by just this one source, we repeat the stacking excluding NGC 4151. In this case we get a maximum TS of 15.1 ($3.5\,\sigma$), corresponding to a best-fit index of $-2.2 \pm 0.4$ and a best-fit flux of $2.0^{+2.0}_{-1.0} \times 10^{-11} \ \mathrm{ph \ cm^{-2} \ {s^{-1}}}$.  

%%%%%%%%%%%%%%%%%%%%%%%%%%%%%%%%%%%%%%%%%%%%%%%
%%%%%%%%%%%%%%%%%%%%%%%%%%%%%%%%%%%%%%%%%%%%%%

%%%%%%%%%%%%%%%%%%%%%%%%%%%%%%%%%%%%%%%%%%%%%%

\subsection{Spectral Energy Distribution of UFOs}

\begin{figure}[t]
\begin{center}
\includegraphics[width=0.49\textwidth]{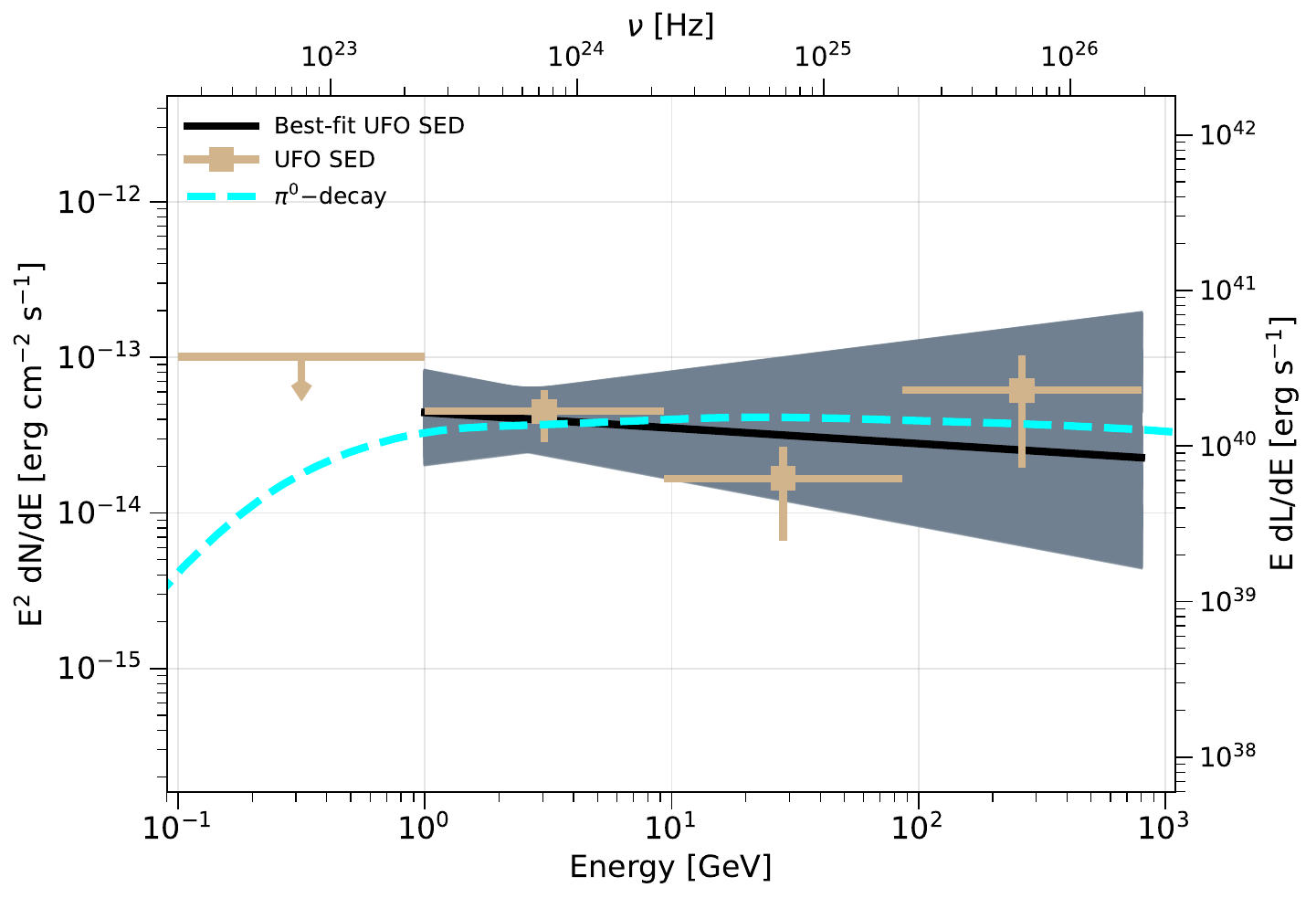} 
\caption{Best-fit UFO SED (black solid line) with 1\,$\sigma$ uncertainty envelope (gray band). The tan data points show the UFO energy flux calculated in four different energy bins.
The dashed cyan line shows our hadronic model (see $\S$~\ref{sec:sed_modeling}), corresponding to an outflow that has propagated to $\sim$20\,pc. The effective redshift $z=0.013$ was used to convert the $\gamma$-ray flux into luminosity.}
\label{fig:ufo_spectrum}
\end{center}
\end{figure}

The best-fit SED for our UFO sample is shown in Figure~\ref{fig:ufo_spectrum}. 
The butterfly plot is constructed by sampling the range of parameter values that are within the 68\,\% confidence contour of the stacked profile. In addition, we calculate the SED flux in 3 logarithmically spaced bins between 1$-$800 GeV. In every bin, we fix the power law index of the UFOs to $-2.0$ and leave all other parameters free to vary. As can be seen, these data points are in agreement with the best-fit SED model. To characterize the UFO spectrum at low energy we repeat the stacking analysis in the energy range 0.1 $-$ 1 GeV, which yields a 95\% flux upper limit ($\Delta \mathrm{logL}=2.71/2$) of 
$5.7 \times 10^{-10} \mathrm{\ ph \ cm^{-2} \ s^{-1}}$. We also overlay our best-fit hadronic model presented in $\S$~\ref{sec:sed_modeling}. 

%%%%%%%%%%%%%%%%%%%%%%%%%%%%%%%%%%%%%%%%%

%%%%%%%%%%%%%%%%%%%%%%%%%%%%%%%%%%%%%%%%%%%%%%
\begin{figure*}[t]
\begin{center}
\includegraphics[width=0.4\textwidth]{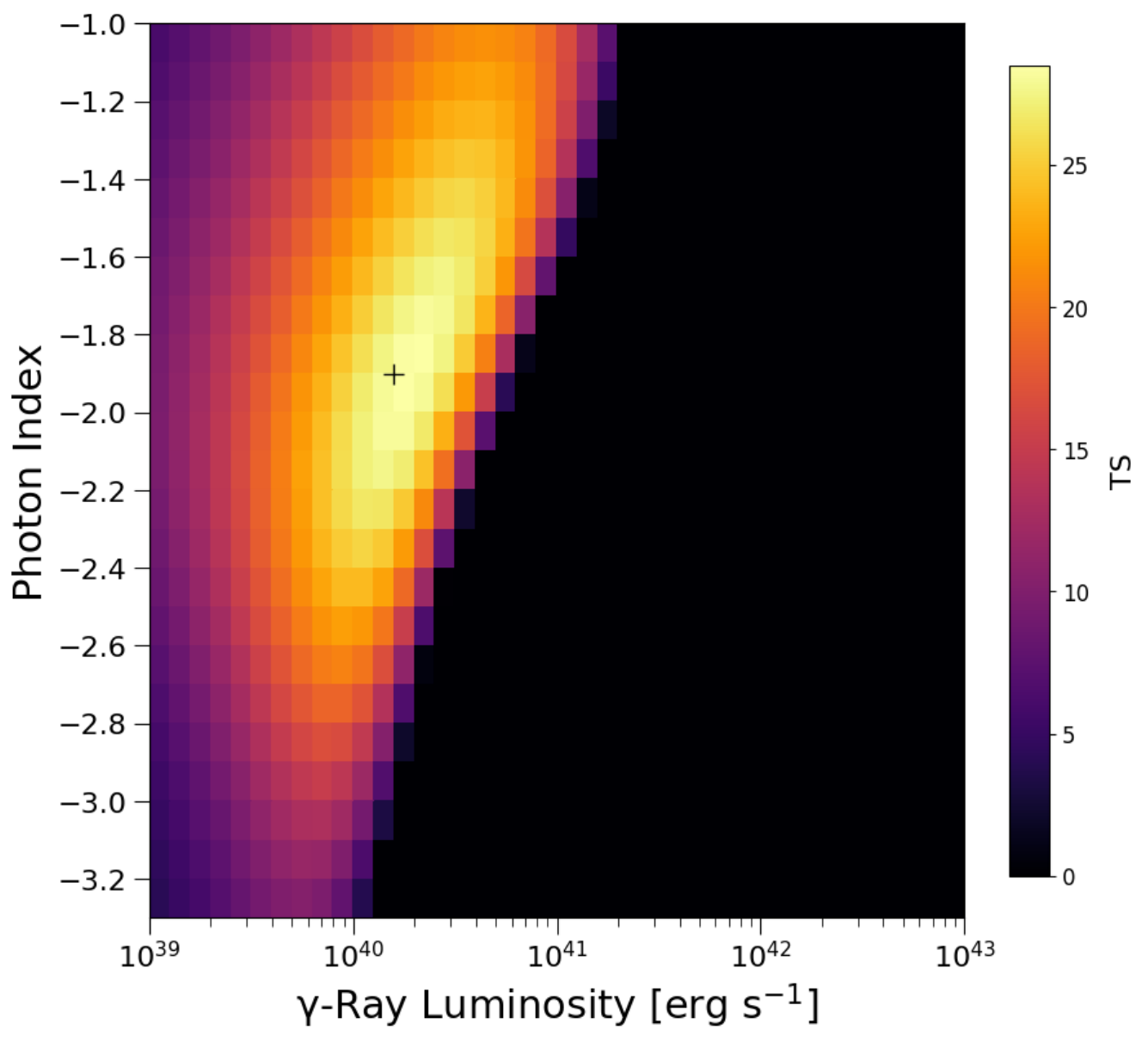} 
\includegraphics[width=0.4\textwidth]{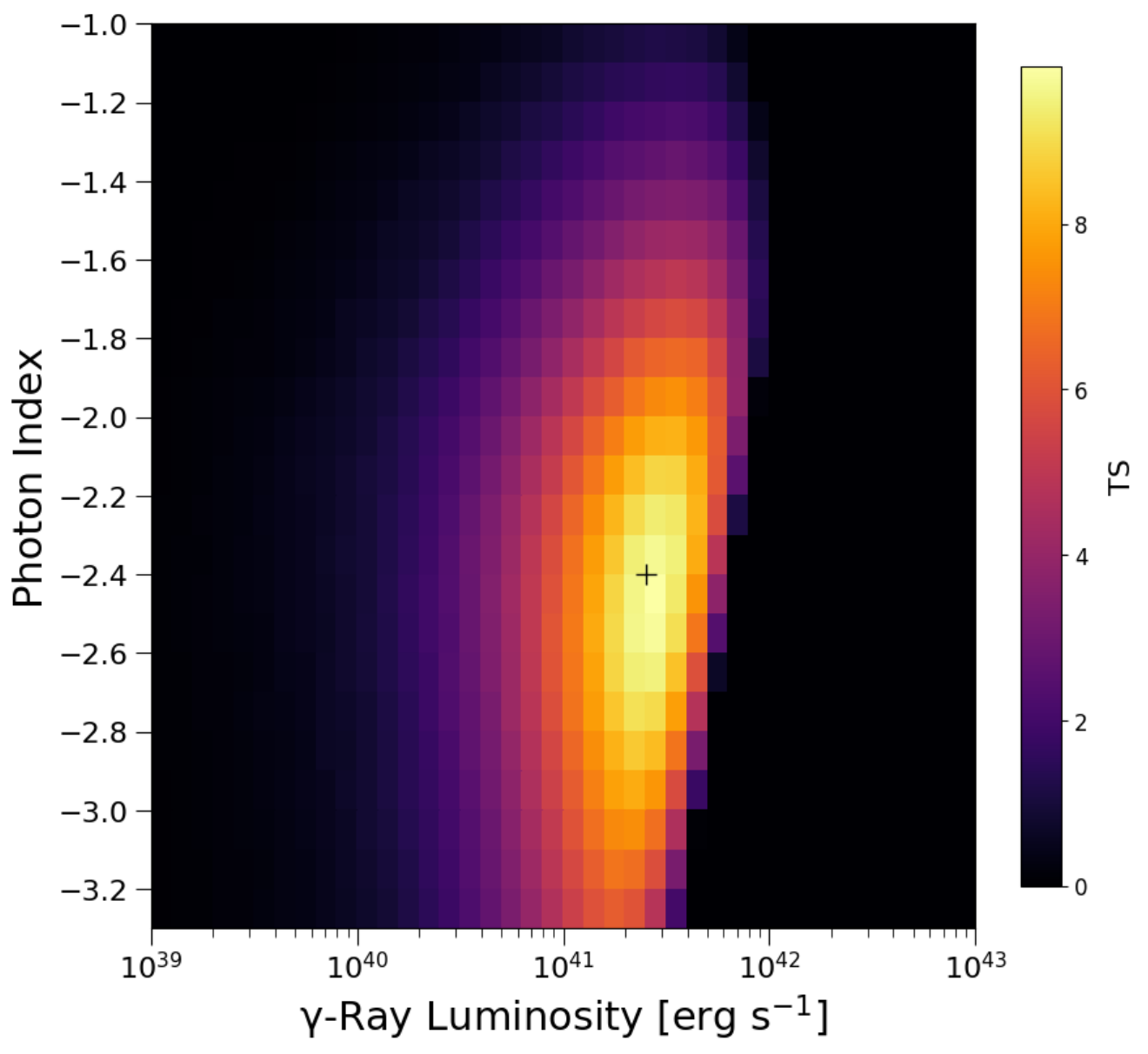}
  \caption{Stacked profiles for bins of  bolometric luminosity (the mean kinetic power bins are also the same). The left and the right panels show the stacking for sources with bolometric luminosity (or kinetic power) below and above the average, respectively.
  The color scale indicates the TS and is set to the maximum value for each bin. The black plus sign gives the best-fit parameters. The first bin consists of 5 sources, with a maximum TS of 28.5 (5.0$\sigma$); and the second bin consists of 6 sources, with a maximum TS of 9.9 ($2.7\sigma$).}
\label{fig:kinetic_bins}
\end{center}
\end{figure*}
%%%%%%%%%%%%%%%%%%%%%%%%%%%%%%%%%%%%%%%%%%%%%%

%%%%%%%%%%%%%%%%%%%%%%%%%%%%%%%%%%%%%%%%%%%%%%
\begin{figure*}[t]
\begin{center}
\includegraphics[width=0.4\textwidth]{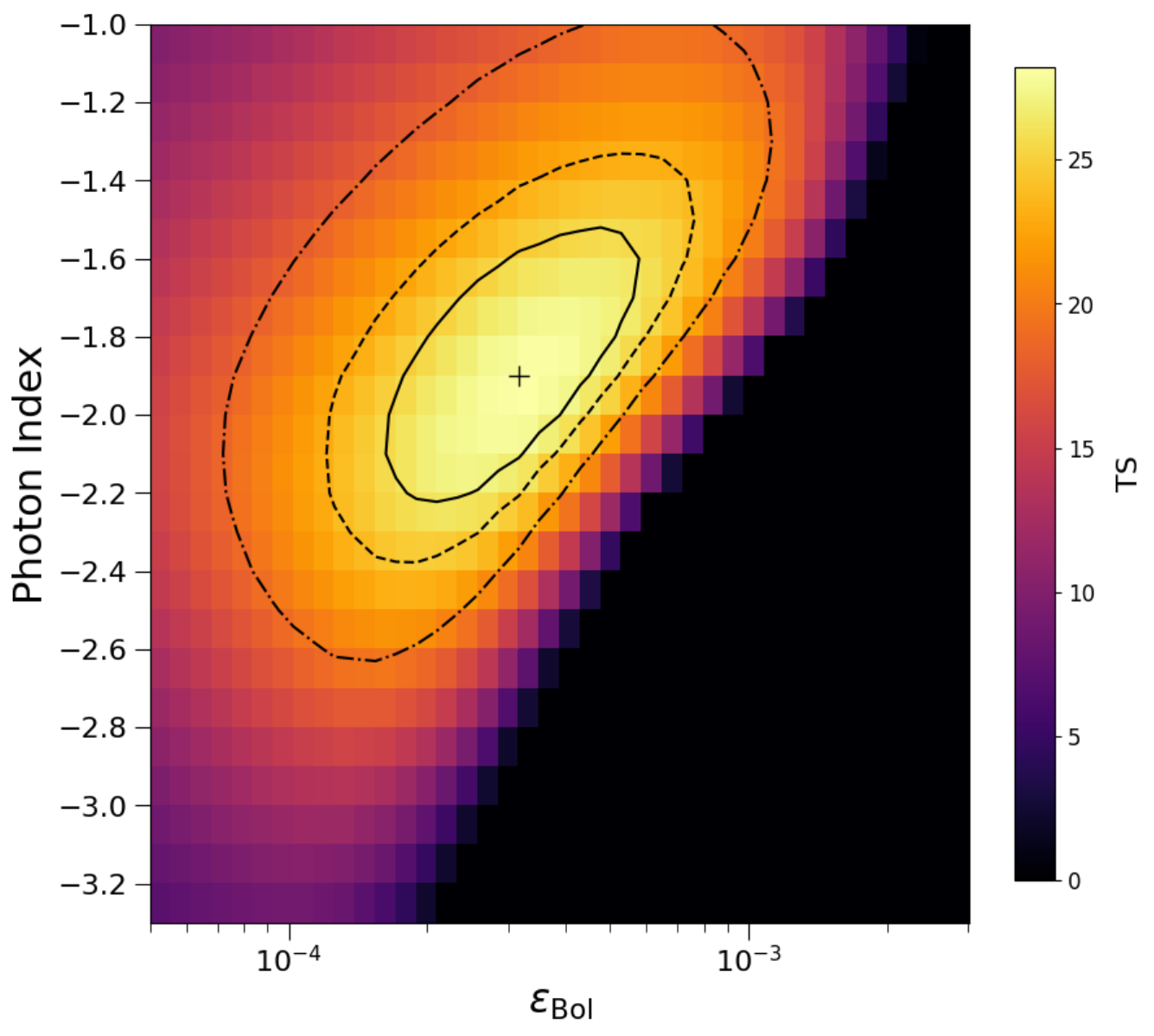} 
\includegraphics[width=0.4\textwidth]{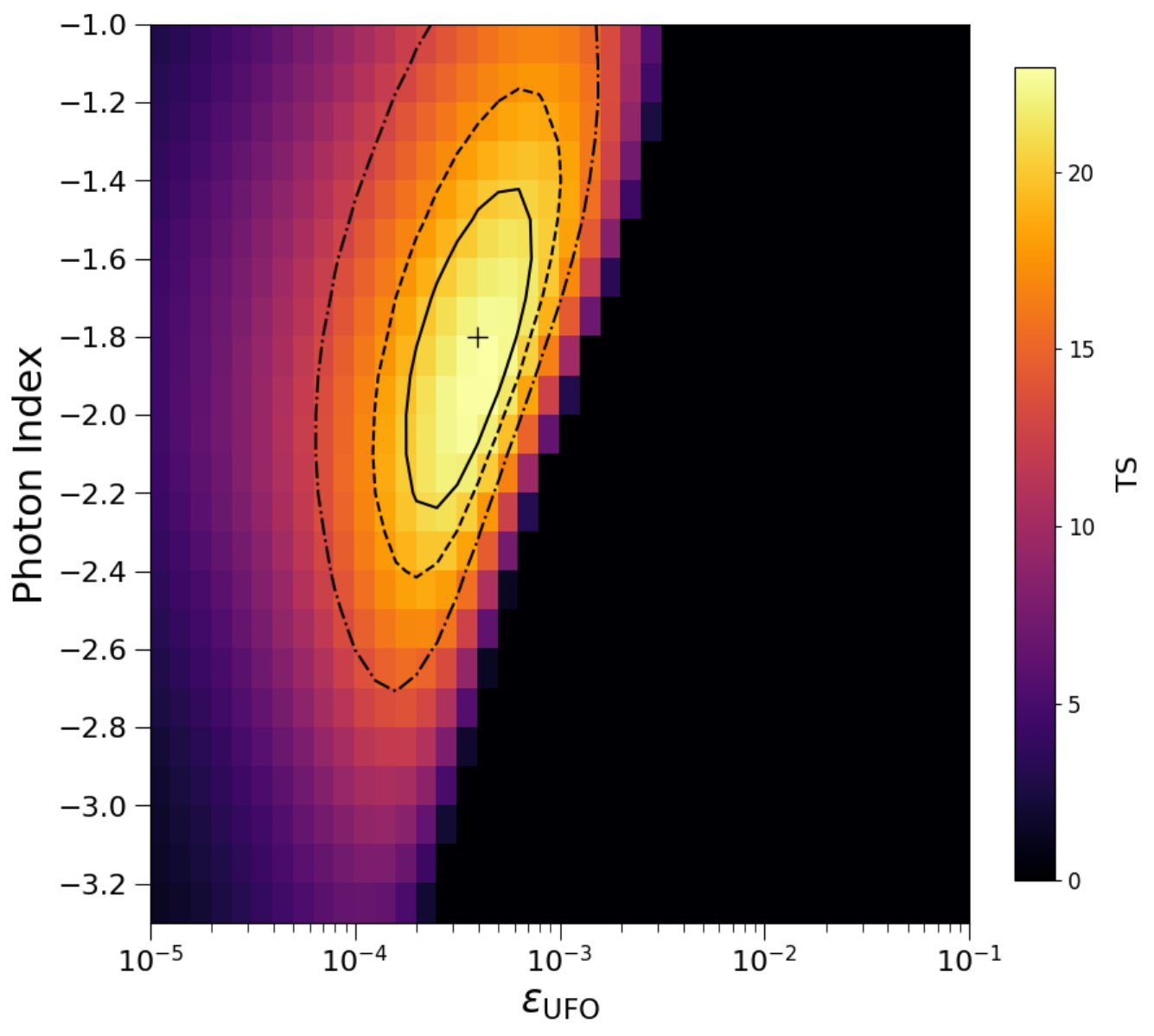}
 \caption{Stacked profiles for bolometric efficiency (left) and kinetic power efficiency (right). The color scale indicates the TS and is set to the maximum value. The black plus sign gives the best-fit parameters. Significance contours (for 2 degrees of freedom) are overlaid on the plot showing the 68\%, 90\%, and 99\% confidence levels, corresponding to $\Delta \mathrm{TS}$ = 2.30, 4.61, and 9.21, respectively.}
\label{fig:efficiency_stack}
\end{center}
\end{figure*}
%%%%%%%%%%%%%%%%%%%%%%%%%%%%%%%%%%%%%%%%%%%%%%

\subsection{Bins of Bolometric Luminosity and Kinetic Power}

We test whether the $\gamma$-ray emission from UFOs scales with AGN bolometric luminosity and outflow kinetic power. To properly take the distance of each source into account, we stack in the luminosity-index space. We take estimates of the bolometric luminosity from the literature, as reported in Table~\ref{tab:ufo_parameters}. Such estimates can be obtained by applying a correction factor to a certain flux, typically the 5100 $\text{\normalfont\AA}$ optical emission,  the 1450 $\text{\normalfont\AA}$ UV emission, or the 2$-$10 keV X-ray emission. Alternatively, the bolometric luminosity can be determined by fitting an SED to the broad-band emission. In any case, the absorption from the host galaxy must be corrected for, which has a large dependence on the viewing angle of the source, and can introduce a rather significant uncertainty. In addition, the contribution from the host galaxy emission also needs to be corrected for {(i.e., UV/IR/Opt emission from the galactic disk)}. Most of the AGN emission is observed in the optical/UV, while $<$10\% is emitted in the X-ray, and thus a broadband SED fitting ensures a more accurate determination of the bolometric luminosity. We therefore search the literature for the most reliable estimates of the bolometric luminosity, and rely on the X-ray determination for only 2 sources (MCG-5-23-16 and SWIFT~J2127.4+5654) for which no other estimates could be found. For sources with multiple estimates we take the geometric mean. The mean of the bolometric luminosity of our sample is $2.5 \times 10^{44} \ \mathrm{erg \ s^{-1}}$, and we create two bins around this value. 

The stacked profiles for the two bins are shown in Figure~\ref{fig:kinetic_bins}. The first bin has 5 sources, with a mean redshift of 0.007. The maximum TS is 28.5 (5.0\,$\sigma$), corresponding to a best-fit index of $-1.9^{+0.3}_{-0.4}$ and a best-fit luminosity of $1.6^{+0.9}_{-0.8} \times 10^{40} \ \mathrm{erg \ {s^{-1}}}$. The second bin has 6 sources, with a mean redshift of 0.04. The maximum TS is 9.9 (2.7\,$\sigma$), corresponding to a best-fit index of $-2.4^{+0.6}_{-0.5}$ and a best-fit luminosity of $2.5^{+1.5}_{-1.5} \times 10^{41} \ \mathrm{erg \  {s^{-1}}}$. The total TS (bin 1 + bin 2) for the stacking in bins is 38.4, compared to 30.1 for the full stack. 

We also stack the $\gamma$-ray luminosity in bins of kinetic power. In general the kinetic power as determined from X-ray observations has a large uncertainty, as can be seen in Table~\ref{tab:ufo_parameters}. Minimum and maximum values are typically reported, corresponding to minimum and maximum radii of the outflow. We  use the geometric mean of the minimum and maximum estimates for our calculations (also incorporating statistical uncertainties in the range). We create two bins around the mean kinetic power, which has a value of $1.8 \times 10^{44} \ \mathrm{erg \ s^{-1}}$. The stacked profiles for the two bins turn out to be the same as for the bolometric bins, as shown in Figure~\ref{fig:kinetic_bins}. 

%%%%%%%%%%%%%%%%%%%%%%%%%%%%%%%%%%%%%%
\begin{figure*}[t!]
\begin{center}
\includegraphics[width=0.47\textwidth]{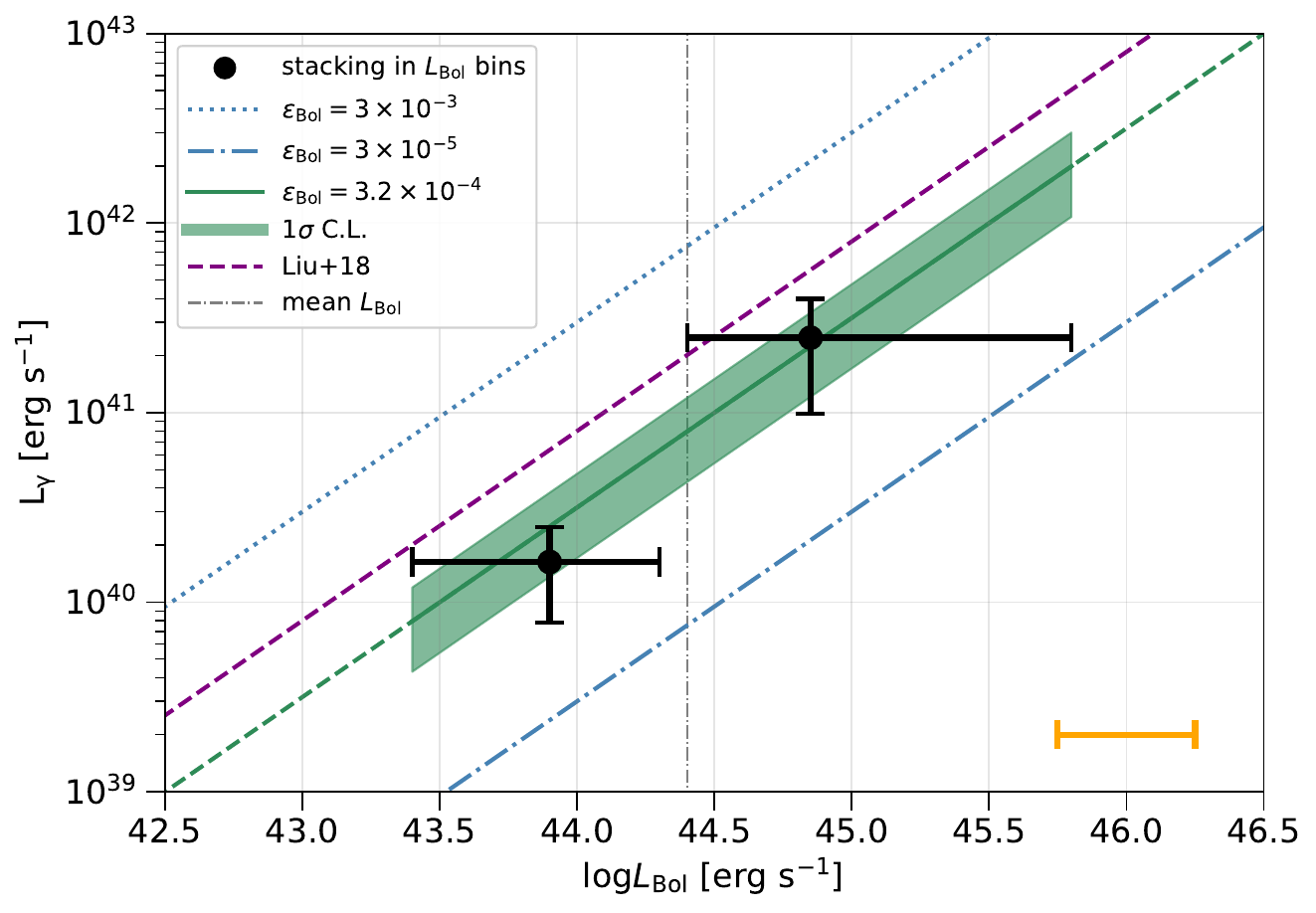} 
\includegraphics[width=0.50\textwidth]{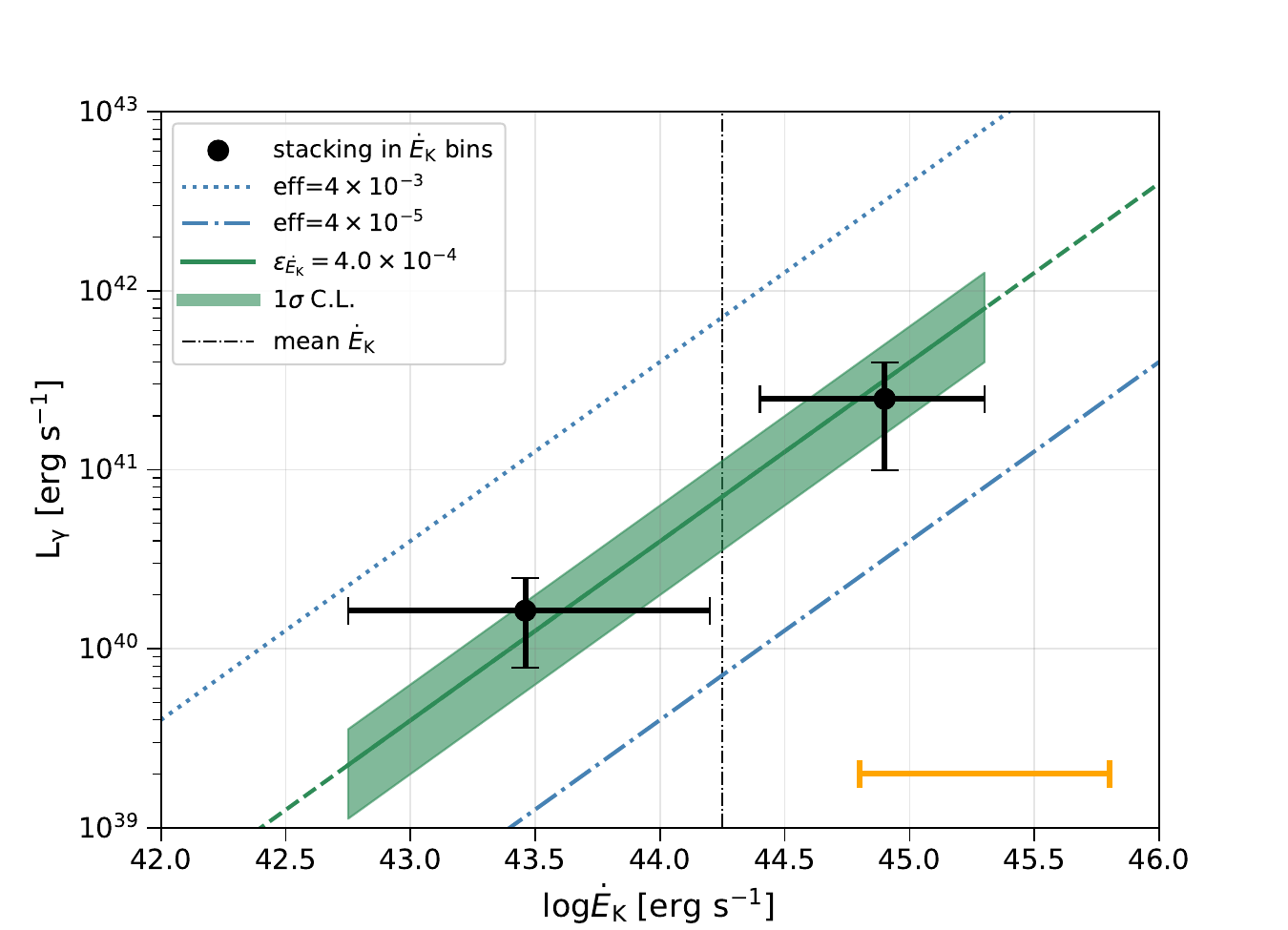} 
\caption{$\gamma$-ray luminosity versus bolometric luminosity (left) and kinetic power (right). The black data points result from stacking in $\gamma$-ray luminosity, and the uncertainty in the x-axis corresponds to the bin widths. The grey dash-dot vertical lines show the value used to divide the bins. The solid green line shows the best-fit resulting from stacking in efficiency, with the green band showing the 1\,$\sigma$ confidence level. For reference, the blue lines show a range of efficiencies within roughly an order of magnitude of the best fit. The 
orange bar in both plots shows the average one-sided uncertainty in individual measurements of AGN bolometric luminosity (left) and kinetic power (right). In the left panel we also overlay the predicted efficiency derived from~\citet[dashed purple line]{Liu+18}. See text for more details.}
\label{fig:ufo_kinetic_power}
\end{center}
\end{figure*}

To further verify the relations found above for the stacking in bins, we perform the stacking analysis using both bolometric efficiency ($\epsilon_{\mathrm{Bol}}=L_\gamma/L_{\mathrm{Bol}}$) and kinetic power efficiency ($\epsilon_{\mathrm{\dot{E}_K}}=L_\gamma/L_{\mathrm{\dot{E}_K}}$). This is done by evaluating for each source the TS of a given $\epsilon_{\mathrm{Bol}}$ (or $\epsilon_{\mathrm{\dot{E}_K}}$) and using that efficiency value, the bolometric luminosity (or kinetic power), and the distance of the source to transform to $\gamma$-ray flux (for a given photon index). Results for these fits are shown in Figure~\ref{fig:efficiency_stack}. The left panel shows the bolometric efficiency, with a best-fit value of $3.2^{+1.6}_{-1.5} \times 10^{-4}$, corresponding to a best-fit index of $-1.9^{+0.3}_{-0.4}$, and a maximum TS of 28.2 (5 $\sigma$). The right panel of Figure~\ref{fig:efficiency_stack}  shows the kinetic power efficiency, with a best-fit value of $4.0^{+2.3}_{-2.0} \times 10^{-4}$, corresponding to a best-fit index of $-1.8^{+0.3}_{-0.4}$, and a maximum TS of 23.0 (4.4 $\sigma$). We note that the best-fit index from the efficiency analysis is slightly harder than the one found by the flux-index stacking, but compatible within 1\,$\sigma$ uncertainties. The small shift observed in the best-fit index value is due to how the TS profiles are weighted differently when stacking in efficiency with respect to flux.

The result for stacking in bolometric luminosity and kinetic power are summarized in Figure~\ref{fig:ufo_kinetic_power}. The left panel shows the $\gamma$-ray luminosity versus bolometric luminosity, and the right panel shows the $\gamma$-ray luminosity versus UFO kinetic power. The black data points are for stacking in bins, and the corresponding best-fit efficiency, along with the 1$\sigma$ confidence interval, is plotted with the green band. Also plotted are lines for different efficiencies under the assumption of a linear scaling. As can be seen, the results on the efficiencies are in very good agreement with the stacking in bins. 

In the left panel of Figure~\ref{fig:ufo_kinetic_power} we also overlay the predicted scaling of $L_{\gamma}$ with $L_{\rm Bol}$  from~\citet{Liu+18}\footnote{Our derivation is made converting the peak 1\,GeV luminosities (reported in their Figure~5) to the 1-800\,GeV energy range using the 
best-fit spectral index of $-2.1$.}. As can be seen,~\citet{Liu+18} predict a nearly linear scaling between the logarithms of the two luminosities (over their $L_{\rm Bol} ({\rm erg s^{-1}})=10^{42}-10^{45}$ range) with an efficiency of $\sim 8 \times 10^{-4}$, which is in reasonably good agreement with the one measured here.

%using the light curves of 1 GeV photons for three different values of $L_{\mathrm{Bol}}$ (see Figure 5 in ~\citet{Liu+18}). We use the peak $\gamma$-ray luminosity for each case and convert it to our energy range of $1-800$ GeV using the best-fit spectral index of 2.1. 

%The purple cross in Figure~\ref{fig:ufo_kinetic_power} shows the resulting value for $L_{\mathrm{Bol}}=10^{45} \ \mathrm{erg \ s^{-1}}$ (the other two data points are outside the plot range). The dashed purple line shows the corresponding efficiency, determined from a least square fit to the three predicted values (which are very close to linear). Remarkably,~\citet{Liu+18} predict an efficiency of $\sim 8 \times 10^{-4}$, which is within a factor of 2.5 to the measured value.} 

\subsection{Representative Luminosity of the Sample}

%%%%%%%%%%%%%%%%%%%%%%%%%%%%%%%%%%%%%%%%%%%%%%

Because the 11 UFO galaxies are detected at fairly different distances, we adopt a weighting scheme to compute the representative luminosity of the sample. In this framework  $\overline{L_{\gamma}}= \frac{\sum_{i=1}^{11} L_{\gamma,i}\times {\rm TS}_i }{{\rm TS}_{tot}}$, where $L_{\gamma,i}$ and $\rm {TS}_i$ are the luminosity and the TS for the $i^{th}$ galaxy at the global best-fit position (1-800\,GeV flux of 2.5$\times10^{-11}$\, ph cm$^{-2}$ s$^{-1}$ and photon index of $-$2.1) and $\rm{ TS}_{tot}=30.1$. The representative luminosity is found to be $\overline{L_{\gamma}}=7.9^{+5.1}_{-2.9}\times 10^{40}$\,erg s$^{-1}$ and would correspond to an effective redshift of $z=0.013$ (adopting the above best-fit parameters). This luminosity is in very good agreement with the one obtained scaling the average bolometric luminosity $L_{\rm Bol}=2.5\times10^{44}$\,erg s$^{-1}$ by the best-fit efficiency ($\epsilon_{\rm Bol}=3.2\times 10^{-4}$). The effective redshift is also very close to the median redshift of the sample ($z=0.013$ vs.~$z=0.014$) making  the TS-weighted luminosity compatible with the median $\gamma$-ray luminosity of the sample.

\subsection{Simulations}

%%%%%%%%%%%%%%%%%%%%%%%%%%%%%%%%%%%%%%%%%%%%%%
\begin{figure}[t]
\begin{center}
\includegraphics[width=0.48\textwidth]{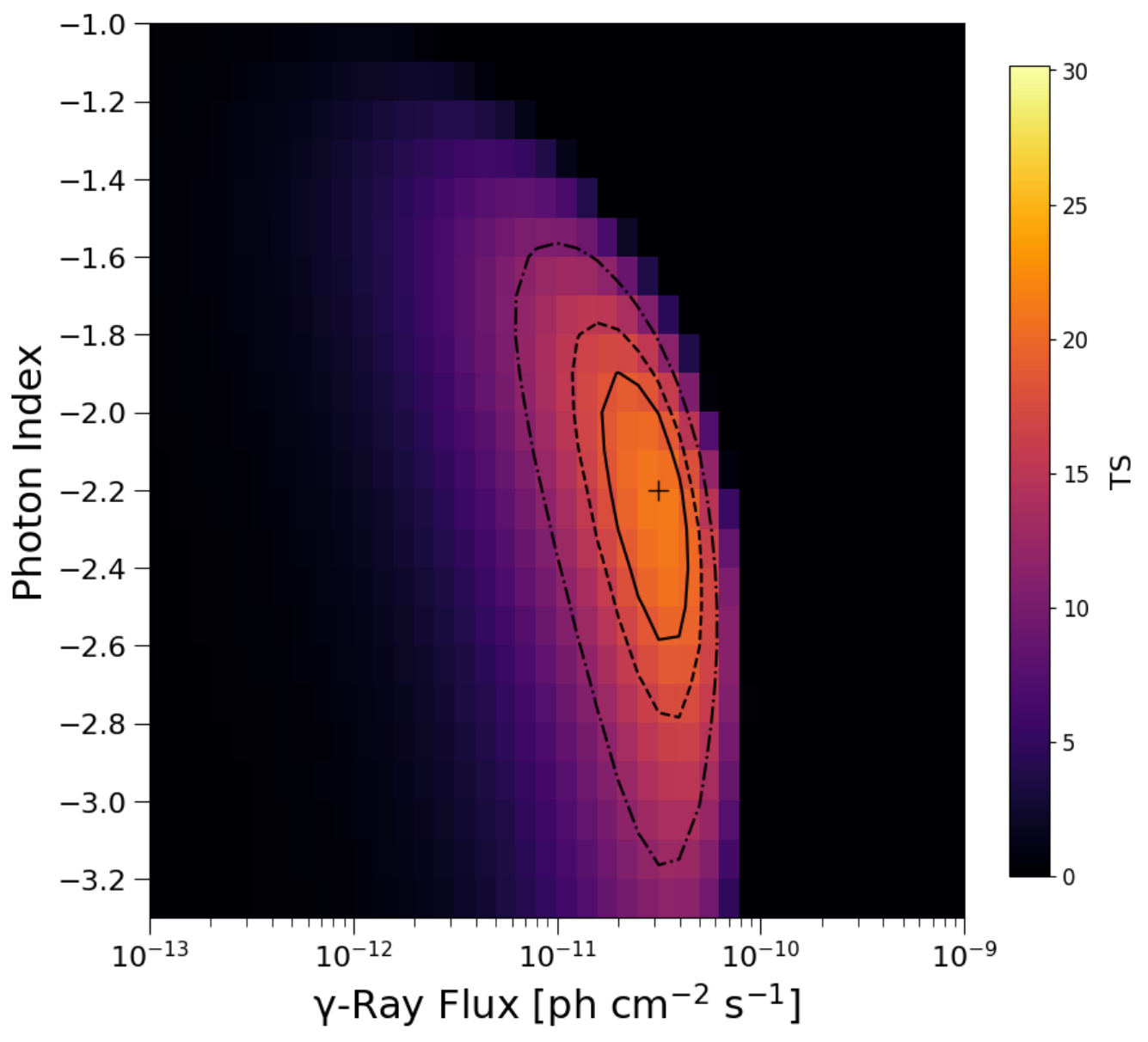} 
\caption{Stacked profile for our simulation run, in which the UFO sources are simulated with an index of $-2.1$ and a flux of $2.5 \times 10^{-11} \mathrm{\ ph \ cm^{-2} \ s^{-1}}$. The color scale indicates the TS, and the plus sign indicates the location of the maximum value, with a $\mathrm{TS} = 21.2 \ (4.2\,\sigma)$. Significance contours (for 2 degrees of freedom) are overlaid on the plot showing the 68\%, 90\%, and 99\% confidence levels, corresponding to $\Delta \ \mathrm{TS}$ = 2.30, 4.61, and 9.21, respectively. The maximum TS of the color scale is set to 30.1 (the maximum value from Figure~\ref{fig:Run_5_JL}).}
\label{fig:Run_7_stacked_profile}
\end{center}
\end{figure}
%%%%%%%%%%%%%%%%%%%%%%%%%%%%%%%%%%%%%%%%%%%%%%

The results presented here are validated using Monte Carlo simulations. We simulate the fields of the 11 UFOs considering the Galactic and isotropic emission (modeled as gll\_iem\_v07 and iso\_P8R3\_SOURCE\_V2\_v1, respectively), background sources from the 4FGL catalog, and our test source at the position of the UFO in each ROI. The UFO spectral parameters are set to be the same as the best-fit values from the data, i.e.~index $=-2.1$ and flux $= 2.5 \times 10^{-11} \ \mathrm{ph \ cm^{-2} \ s^{-1}}$. For simplicity we use the standard event type (evtype$=3$), i.e.~we do not use the four different PSF event types. The data is simulated using the \textit{simulate\_roi} function from Fermipy. The simulation is created by generating an array of Poisson random numbers, where the expectation values are drawn from the model cube\footnote{More information on generating the simulations is available at \url{https://fermipy.readthedocs.io/en/latest/fermipy.html}.}. 
Finally, we run our stacking pipeline on the simulated data. We recover the input values, with a best-fit index of $-2.2^{+0.4}_{-0.2}$, a best-fit flux of $3.2^{+1.8}_{-1.6} \times 10^{-11} \ \mathrm{ph \ cm^{-2} \ {s^{-1}}}$, and a maximum TS of 21.2 ($4.2\,\sigma$). The stacked profile is shown in Figure~\ref{fig:Run_7_stacked_profile}. Overall the results from the simulation are consistent with the real data. 

\section{Additional Tests}
\label{sec:additional_tests}
\subsection{Control Sample}

\begin{figure}[t]
\begin{center}
\includegraphics[width=0.47\textwidth]{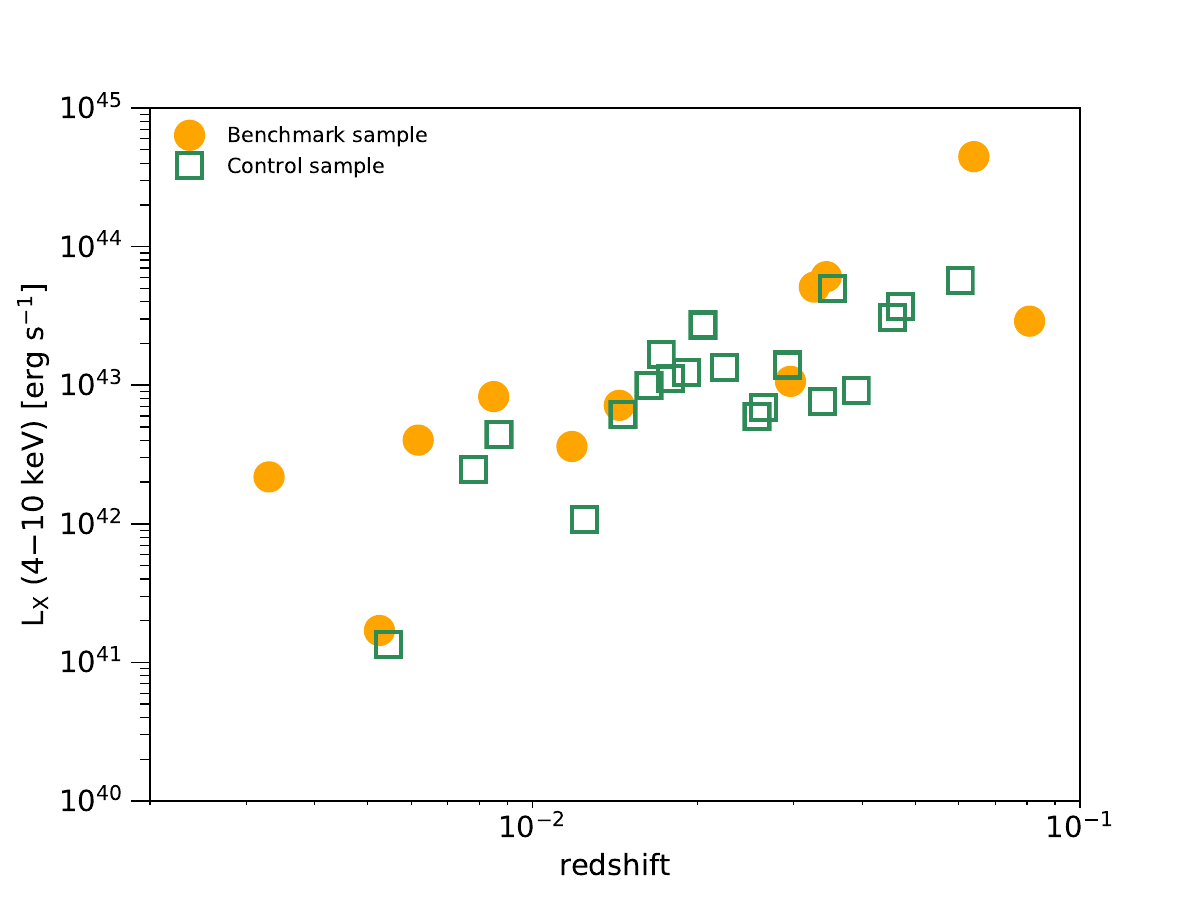} 
\caption{Comparison of redshift and X-ray luminosity (4$-$10 keV) for the control sample and benchmark sample, as indicated in the legend.}
\label{fig:control_benchmark_comparison}
\end{center}
\end{figure}

We repeat the analysis with a sample of 20 low redshift ($z<0.1$) radio-quiet AGN that do not have UFOs. The sources were selected from the samples of~\citet{tombesi2010evidence} and~\citet{2020MNRAS.493.1088I} for which no UFO was found. The sample of~\citet{tombesi2010evidence} is based on absorption features, while the sample of~\citet{2020MNRAS.493.1088I} uses the excess variance method. Of the 20 sources in our control sample, there are 10 sources in common between the two studies, 4 additional sources from~\citet{tombesi2010evidence}, and 6 additional sources from~\citet{2020MNRAS.493.1088I}. 
For reference, the list of sources in the control sample is given in Table~\ref{tab:control_sample}. Figure~\ref{fig:control_benchmark_comparison} shows  that the benchmark and control samples are well matched in X-ray luminosity and redshift.

\begin{figure}[t]
\begin{center}
\includegraphics[width=0.45\textwidth]{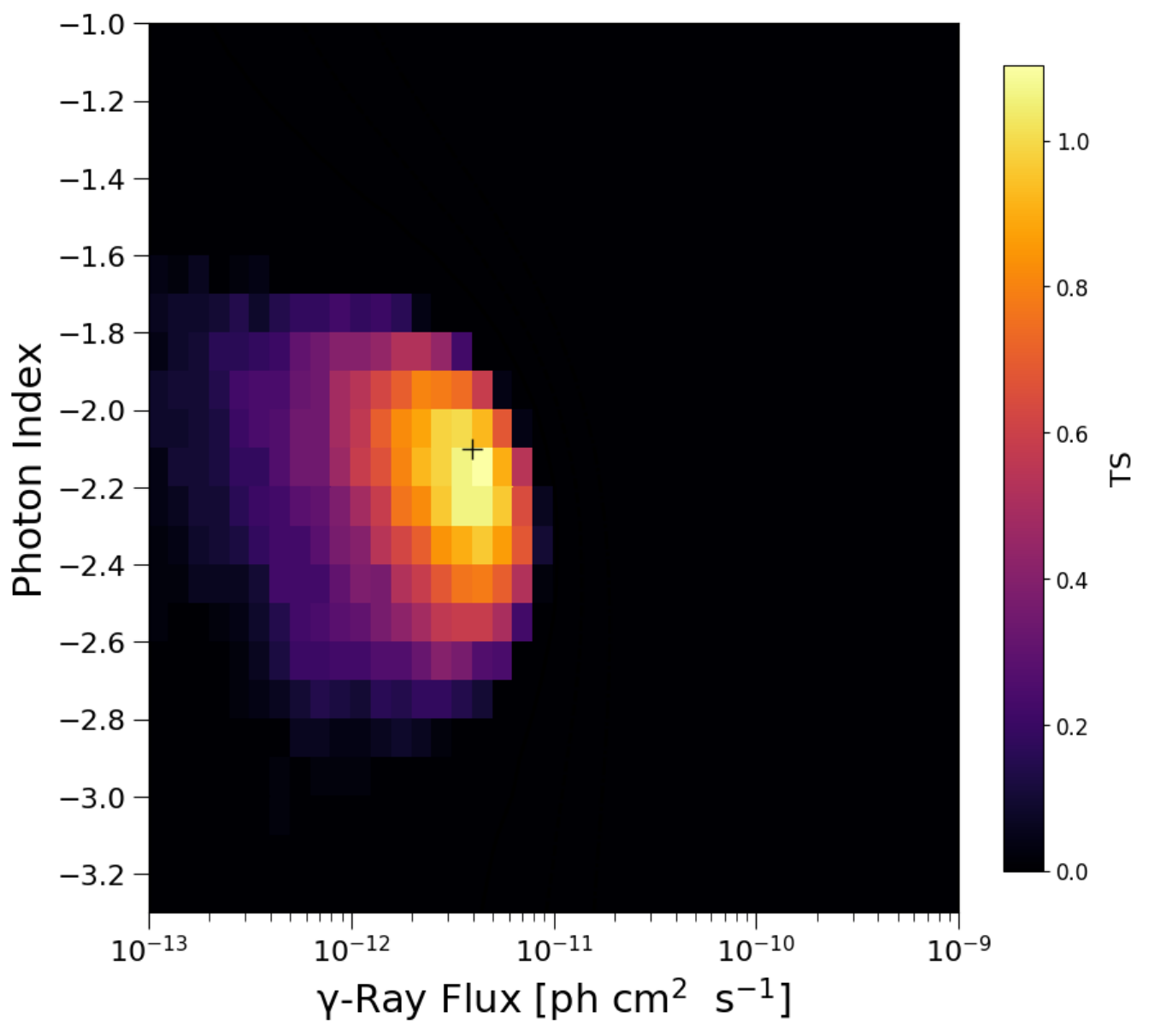} 
\caption{Stacked profile for our control sample consisting of 20 nearby ($z<0.1$) radio-quiet AGN with no UFOs (i.e.~a UFO has been searched for but none has been detected). No signal is detected, with a maximum TS of 1.1.}
\label{fig:control_sample}
\end{center}
\end{figure}

Results for the stacked profile are shown in Figure~\ref{fig:control_sample}. No signal is detected, with a maximum TS of 1.1. Using the profile likelihood method and a photon index of $-$2.0, the upper limit on the flux (1$-$800 GeV) at the 95\% confidence level is $8.8 \times 10^{-12} \mathrm{\ ph \ cm^{-2} \ s^{-1}}$. This supports the interpretation of the $\gamma$-ray emission being due to the outflow rather than other processes in AGN.

\begin{deluxetable}{lccccc}
\tabletypesize{\scriptsize}
\tablecaption{Control Sample}\label{tab:control_sample}
\tablewidth{0pt}
\tablehead{
\colhead{Name} & \colhead{RA} & \colhead{DEC} & \colhead{Redshift} & \colhead{IR Lumin.} &\colhead{1.4\,GHz flux}  \\
\colhead{} & \colhead{} & \colhead{} & \colhead{} & \colhead{ log (L$_{\odot}$)} & \colhead{[mJy]} 
}
\decimalcolnumbers
\startdata
ESO 198-G024    &39.58       &-52.19 & 0.046 & \nodata& \nodata\\
Fairall 9       &20.94       &-58.81 &0.047 & \nodata& \nodata\\
H 0557-385      &89.51       &-38.33 & 0.034 & \nodata& \nodata\\
MCG+8-11-11     &88.72       &46.44  & 0.020 & 11.1 & 286\\
Mrk 590         &33.64       &-0.77  & 0.026 & \nodata & \nodata\\
Mrk 704         &139.61      &16.31  & 0.029 & \nodata & \nodata\\
NGC 526A        &20.98       &-35.07  &0.019  & 10.5 & 13.9\\
NGC 5548        &214.50      &25.14  &0.017  & \nodata & \nodata\\
NGC 7172        &330.51      &-31.87  &0.0090  & 10.4 & 37.6\\
NGC 7469        &345.82      &8.874  & 0.016  & 11.6 & 181 \\
ESO 113-G010    &16.32       &-58.44  &0.027  & \nodata& \nodata\\
ESO 362-G18     &79.90       &-32.66  &0.012  & \nodata& \nodata\\
IRAS 17020+4544 &255.88      &45.68  &0.060  & 11.6 & 129\\
MS22549-3712    &344.41      &-36.94  &0.039  & \nodata& \nodata\\
NGC 1365        &53.40       &-36.14  &0.0055  & 10.9 & 534\\
NGC 4748        &193.05      &-13.41  &0.015  & 10.4 & 14.3\\
Mrk 110         &141.30      &52.29  &0.035  & \nodata& \nodata\\
IRAS 05078+1626  &77.69       &16.50  &0.018  & 10.8 & 6.3\\
ESO 511-G30     &214.84      &-26.64  &0.022  & \nodata& \nodata\\
NGC 2110        &88.05       &-7.46 &0.0078  & 10.3 & 300\\
\enddata
\tablecomments{See~\citet{tombesi2010evidence} and~\citet{2020MNRAS.493.1088I} for further details of the sources.
 The IR luminosity is reported in the 8-1000\,$\mu$m range and derived from IRAS \citep{kleinmann86,moshir90}. The radio fluxes are derived from NVSS \citep{condon_1998}.
}
\end{deluxetable}.
%%%%%%%%%%%%%%%%%%%%%%%%%%%%%%%%%%%%%%%%%%%%%%
%%%%%%%%%%%%%%%%%%%%%%%%%%%%%%%%%%%%%%%%%%%%%%

%%%%%%%%%%%%%%%%%%%%%%%%%%%%%%%%%%%%%%%%%%%%%%
%%%%%%%%%%%%%%%%%%%%%%%%%%%%%%%%%%%%%%%%%%%%%%

\subsection{Alternative UFO Samples}

The fractional excess variance method was recently used in~\citet{2020MNRAS.493.1088I} to search for UFOs in the samples of \citet{tombesi2010evidence} and~\citet{2016MNRAS.462..511K}. Overall, the results are in agreement with the past literature, finding that UFOs are a relatively widely observed phenomena in nearby AGN. However, there are differences with respect to previous studies in regards to which sources are classified as UFOs, and the corresponding UFO parameters.

As the authors mention in~\citet{2020MNRAS.493.1088I}, their method  relies on the variability of the strength of the emission 
(or absorption) features and is less sensitive in detecting cases where these features may vary in energy. The excess variance method is well suited for detecting UFOs in objects that show small changes in the energy of the UFO, but large changes of the equivalent width for the same energy. This is one reason why the excess variance method can potentially miss objects that were detected in spectral-timing analyses that model individual spectra in single epochs.
 
As an additional a-posteriori test we perform our stacking analysis with the UFO sample determined in~\citet{2020MNRAS.493.1088I},
relying on sources classified as either likely outflows or possible outflows therein. Additionally, we use the same selection criterion as for our benchmark sample, i.e.~$z<0.1$ and $v>0.1c$. This gives a sample of 18 sources. The maximum TS is 13.0 (3.2$\sigma$), corresponding to a best-fit flux of $\sim 2.0 \times 10^{-11} \ \mathrm{ph \ cm^{-2} \ {s^{-1}}}$ and a best-fit index of $\sim -2.4$. These results, although less significant, are in good agreement with those from our benchmark sample and show that there is $\gamma$-ray emission associated to UFOs independently of how these sources were selected.

\subsection{Emission from Star-formation activity}

Star-forming galaxies are known $\gamma-$ray emitters because of their CR population, which is accelerated at the shock fronts of supernova remnants and pulsar wind nebulae~\citep{ajello2020}. The ensuing $\gamma$-ray emission is known to correlate well with the total infrared (IR)  luminosity (8-1000\,$\mu$m), which is a tracer of star formation. 

We find that the average total IR luminosity is $\log (L_{\odot})=10.4$ (see Table~\ref{tab:velocity_dispersion}). According to the correlation reported in~\citet{ajello2020} this implies an average $\gamma$-ray luminosity ($>$1\,GeV) of 2.2$\times10^{39}$\,erg s$^{-1}$. This is about 40 times smaller than the observed luminosity and implies that the contamination due to star-formation activity to the signal observed in the UFO sample is negligible.

As an additional test we searched for IR fluxes for the galaxies in the control sample (see Table~\ref{tab:control_sample}). We could find data for nine galaxies with an average total  IR luminosity of $\log (L_{\odot})=10.8$ (compared to 10.4 for the benchmark sample). The stacking of this subset of galaxies in the control sample yields no detection (TS=0.04 and 95\% flux UL=$1.1\times 10^{-11} \ \mathrm{ph \ cm^{-2} \ s^{-1}}$) confirming that the contamination of the signal due to star formation is negligible.

\subsection{Emission from Potential Jets in Radio-quiet AGN}

The vast majority of the $\gamma$-ray sources detected by the LAT are powered by relativistic jets closely aligned to the line of sight~\citep{4LAC}. Some of the sources in our sample, particularly NGC~4151, may have a jet. However, there are several reasons why the $\gamma$-ray emission that we observe is unlikely to be produced by the jets, which may be present in these radio-quiet AGN. The best-studied system\footnote{Other sources like NGC~5506 and NGC~7582 do not have resolved radio jets down to 0.1$''$, while MCG-5-23-16 has a resolved morphology suggesting the presence of a jet~\citep{orienti2010}.}
is NGC~4151, for which an elongated series of knots, possibly associated with a jet, have been detected in radio~\citep{johnston82,wilson82}. This jet has an angle of $\approx40^{\circ}$ with respect to the line of sight and a speed $\approx$0.04$c$~\citep{williams2017}. This is among the lowest speeds measured for a jet and indicates non-relativistic  motion, likely due to thermal plasma~\citep{ulvestad2005}. NGC~4151's jet lies on the opposite end of the spectrum of jets detected by the LAT, which are aligned often within 1$^{\circ}-2^{\circ}$~\citep{pushkarev2017}, highly relativistic~\citep{lister2016}, dominated by non-thermal emission, and found only in radio-loud AGN~\citep{4LAC}.

Moreover, the emission from jets is not expected to correlate with the bolometric luminosity of radio-quiet AGN or the outflow kinetic power. It should also be noted that the sources in our sample follow the $L_{{\rm 22\ GHz}}/L_{\rm 14-195\, keV}\sim10^{-5}$ trend indicating a contribution to the radio luminosity from the hot AGN corona~\citep{smith2020}. 
Finally, the analysis of winds and jets in a sample of radio-loud AGN
provides evidence for a wind-jet bimodality, where winds are the strongest when jets are the weakest \citep[as measured by the radio-loudness parameter][]{mehdipour2019}.

More importantly, the same 9 galaxies in the control sample for which we could find IR data also have 1.4\,GHz fluxes (see Table~\ref{tab:control_sample}). This sample is well matched in terms of radio fluxes and redshift to our benchmark sample and as reported above yields no $\gamma$-ray detection.

\section{SED Modeling}
\label{sec:sed_modeling}

We assume as in~\citet{wang2016contribution} and~\citet{lamastra2017extragalactic}, that the $\gamma$-ray emission is dominated by hadronic processes resulting from diffusive shock acceleration (DSA). 
In order to model these processes in detail, we first calculate proton distributions using the Cosmic Ray Analytical Fast Tool (CRAFT), a code that uses a semi-analytical formalism for DSA described in~\citep{blasi02, amato+06,caprioli+10b} and references therein. CRAFT self-consistently solves the diffusion-convection equation (e.g.,~\citet{skilling75a}) for the transport of non-thermal particles in a quasi-parallel, non-relativistic shock, including the dynamical effects of both accelerated particles and the magnetic turbulence they generate~\citet{Damiano:2011tp,caprioli12}. CRAFT also uses microphysical information (particle injection, diffusion, magnetic field amplification) tuned on self-consistent kinetic plasma simulations of non-relativistic shocks~\citep{caprioli+14a,caprioli+14b,caprioli+14c,caprioli+15,haggerty+19a}. Thus, given basic information about UFO shock hydrodynamics (age, velocity, and ambient density), CRAFT self-consistently predicts an instantaneous proton distribution.

\begin{figure}[t]
    \centering
    \includegraphics[width=0.48\textwidth]{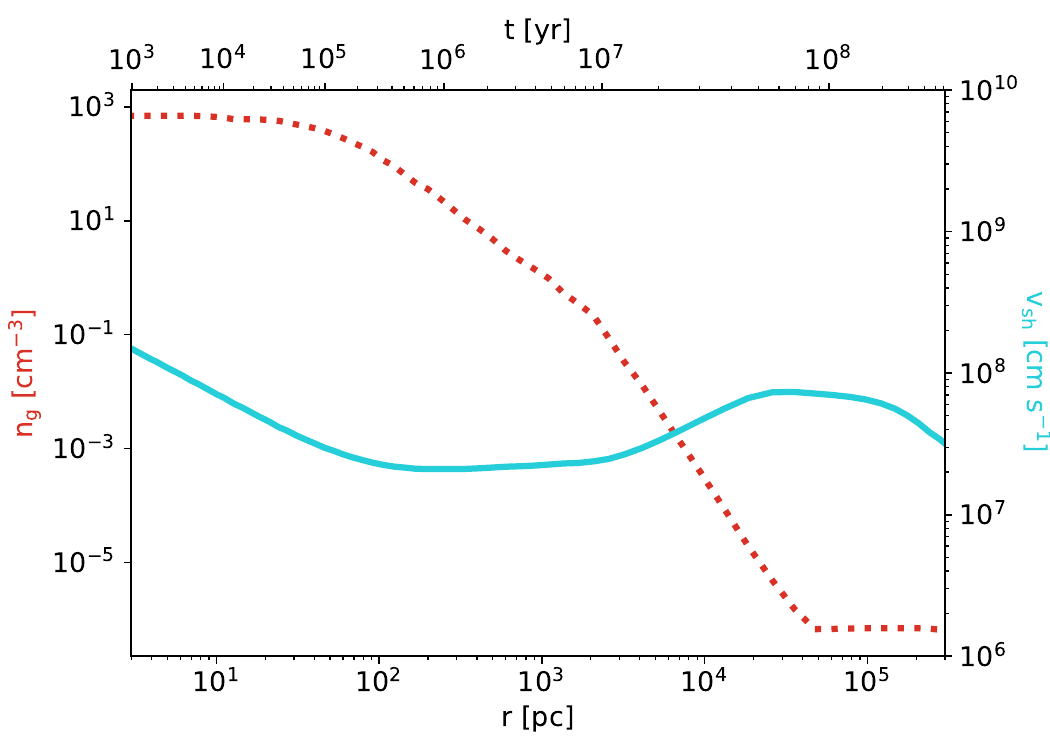}
    \caption{Ambient gas density (red dotted line) and velocity (blue solid line) profiles used in our UFO model. Both profiles come  from the models calculated in~\citet{Liu+18}, for an AGN bolometric luminosity of  $L_{\rm Bol}=2.5\times10^{44}$\,erg s$^{-1}$.}
    \label{fig:hydro_liu}
\end{figure}

To model the cumulative photon distribution of a UFO, we use the hydrodynamic model for the forward shock evolution calculated in~\citet{Liu+18} and shown in Figure~\ref{fig:hydro_liu}. More specifically, \cite{Liu+18} calculate the forward shock evolution \citep[as in][and elsewhere in the literature]{lamastra2017extragalactic, wang2016contribution, wang2016b} using the thin-shell approximation, in which a spherically symmetric shell of negligible thickness expands due the pressure of a hot bubble inside it.  \cite{Liu+18} adopts a broken power law density profile for the ambient gas, $\propto R^{-2}$ inside the disk radius and $\propto R^{-3.95}$ outside the disk. However, \cite{Liu+18} also includes a flat core in the inner 100\,pc of the galaxy to prevent high central densities that are inconsistent with observations, as well as a constant density beyond the virial radius of the galaxy to account for the presence of the intergalactic medium. This profile reproduces well the stellar velocity dispersion in the bulge of the galaxies in our sample (see Figure~\ref{fig:mass_disp}). 
% It gives 180km/s
Both the forward shock evolution and density profile apply to the case of an AGN with a  bolometric luminosity of $L_{\rm Bol}=2.5\times10^{44}$\,erg s$^{-1}$ (consistent with our measurement) and are both shown in Figure \ref{fig:hydro_liu}. Of course, the use of a 1D model has limitations; it cannot account for a more complex ambient medium meaning that inferred values such as the forward shock age and radius are only approximate. However, given that the model in \cite{Liu+18} yields $\gamma$-ray spectra in good agreement with observations, this calculation demonstrates that the $\gamma$-ray emission reported in this work can be explained by a population of UFOs with reasonable parameters.

%For consistency, we assume that the density profile for the representative galaxy is also that used in~\citet{Liu+18}. The density profile reproduces well the stellar velocity dispersion in the bulge of the galaxies in our sample (see Figure~\ref{fig:mass_disp}). Both the forward shock evolution and density profile apply to the case of an AGN with a  bolometric luminosity of $L_{\rm Bol}=2.5\times10^{44}$\,erg s$^{-1}$ (consistent with our measurement) and are both shown in Figure~\ref{fig:hydro_liu}. 

%%%%%%%%%%%%%%%%%%%%%%%%%%%%%%%%%%%

%%%%%%%%%%%%%%%%%%%%%%%%%%%%%%%%%%
\begin{figure*}
    \centering
    \includegraphics[width=0.49\textwidth]{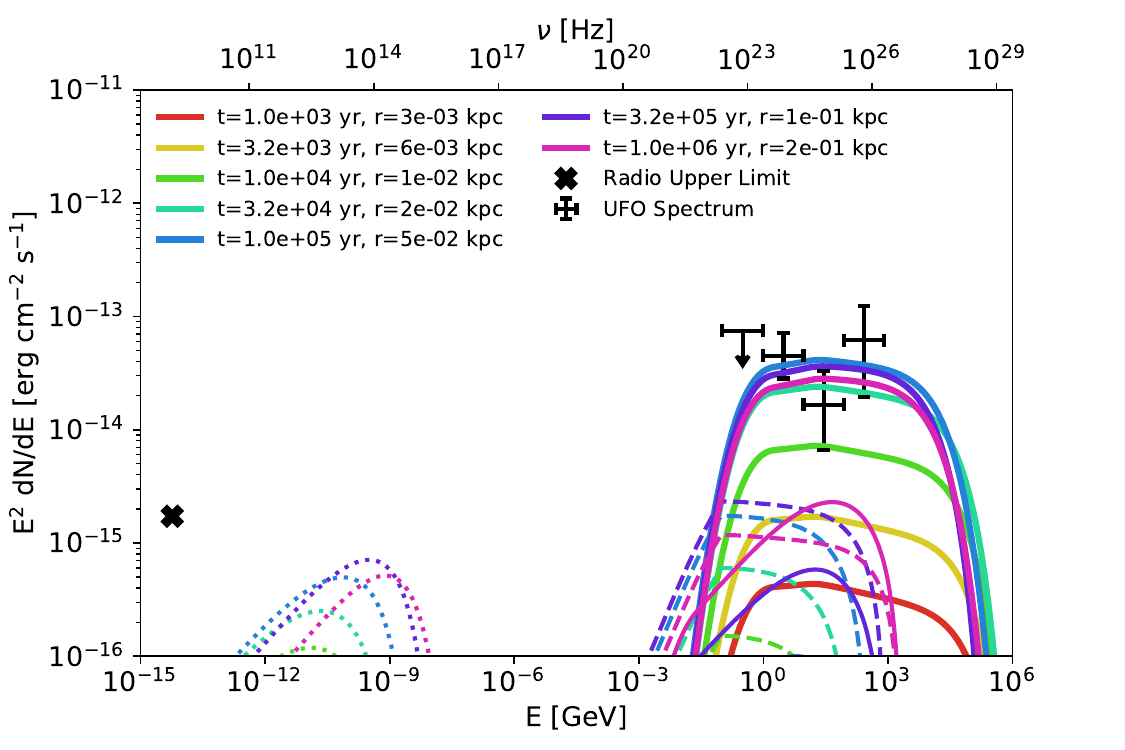}
    \includegraphics[width=0.49\textwidth]{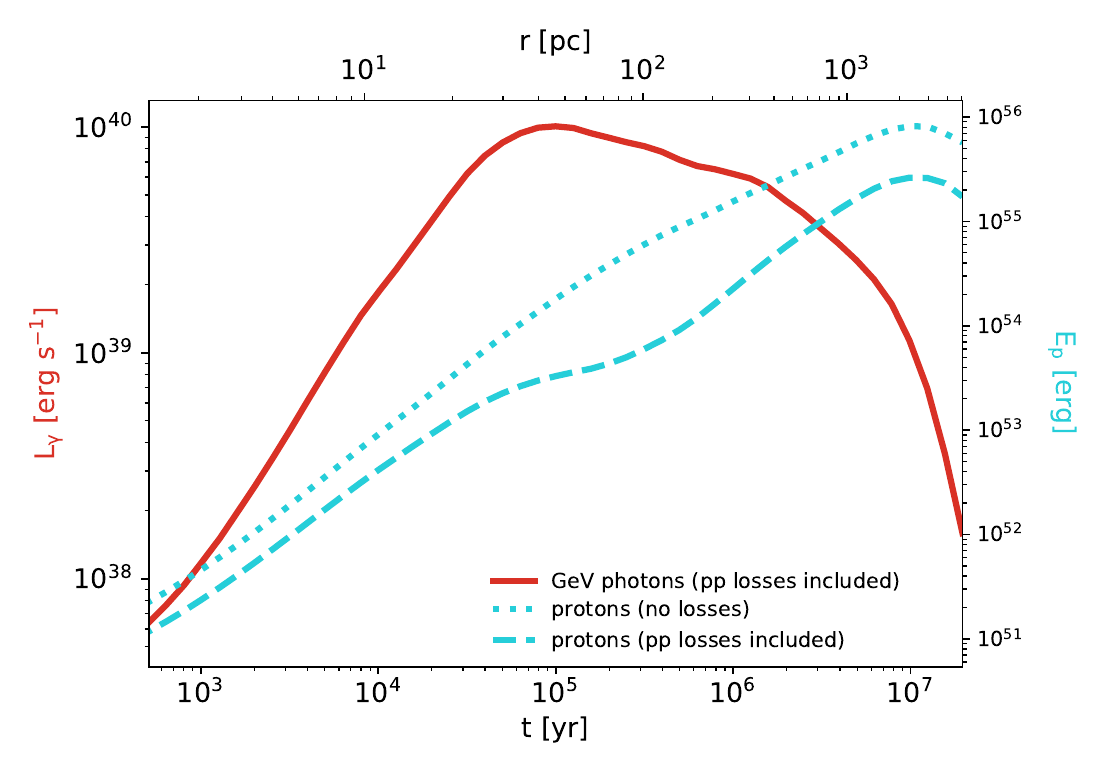}
    \caption{\textbf{Left:} Predicted multiwavelength SED of the UFO's nonthermal emission as a function of time. Synchrotron emission (dotted curves), bremsstrahlung emission (dashed curves), inverse-Compton emission (thin solid curves) and emission from $\pi^0$-decay (thick solid curves) are shown. The inverse-Compton emission remains subdominant despite assuming an artificially enhanced stellar radiation field of energy density 100 eV cm$^{-3}$. Also overlaid is the observed $\gamma$-ray flux as shown in Figure~\ref{fig:ufo_spectrum} and the average radio upper limit from Table~\ref{tab:velocity_dispersion}. Note that the leptonic emission produced at early times often does not appear as it falls below the plot range. \textbf{Right:} Light curve of a UFO-powered forward shock moving through a representative galaxy. The total energy in CRs is shown before and after proton-proton losses are included (blue dotted and dashed lines, respectively), as is the $\gamma$-ray luminosity at 1 GeV (red solid line).}
    \label{fig:ufo_spectrum_full}
\end{figure*}

After using CRAFT to calculate the proton distribution produced at each time step of the shock evolution (see Figure~\ref{fig:hydro_liu}), the resulting instantaneous distributions are weighted and shifted in energy to account for adiabatic losses as in~\citet{diesing+19}. Energy losses due to proton-proton collisions--which are eventually responsible for the UFO's $\gamma$-ray emission--are also taken into account by calculating the collision rate for each distribution at each time step, assuming a target proton density given by the adiabatically expanded post-shock density of a given shell. We further assume that a proton loses half its energy in a single collision (i.e., we assume an inelasticity $\kappa = 0.5$, as in~\citet{Liu+18}). In other words, the accelerated proton population is treated as a series of adiabatically expanding shells, with the outermost shell located at the forward shock. Each of these shells experience proton-proton collisions--and by extension, produce $\gamma$-rays at every time step. Thus, to calculate a UFO's $\gamma$-ray spectrum at a given time, we simply take these weighted proton distributions and convert them to photon spectra using the radiative processes code \verb|naima|~\citep{naima}. We then add these photon spectra together to produce a cumulative SED. Note that Coulomb losses are neglected in this calculation, as they are subdominant for protons with energies $\gtrsim 1$ GeV~\citep{Mannheim+94}.

The result is an estimate of a UFO's SED at every stage of its evolution, as shown in the left panel of Figure~\ref{fig:ufo_spectrum_full}. We obtain $\gamma$-ray luminosities consistent with those calculated in~\citet{Liu+18} and find that the observed $\gamma$-ray emission can be explained by a forward shock that has traveled a distance between $0.02-0.3$ kpc from the SMBH (age of $t=(0.3-10)\times10^5$ yr). The modeled CR and $\gamma$-ray light curves of the UFO are also shown in the right panel of Figure~\ref{fig:ufo_spectrum_full}. It is worth noting that the total energy in CRs--and thus the UFO's $\gamma$-ray luminosity --naturally cuts off after roughly ten million years due to the fact that the ambient density in the reference galaxy decreases substantially with radius, thereby reducing the available energy flux across the shock. 

We also estimate instantaneous electron distributions from our instantaneous proton distributions by using the formalism in~\citet{zirakashvili+07} and accounting for the effects of both adiabatic and synchrotron losses in our weighting \citep[see][]{diesing+19}. To confirm that the UFO's synchrotron emission remains below the average radio upper limit from Table~\ref{tab:velocity_dispersion}, the relative normalization of these electron distributions is taken to be a factor of a few larger than that needed to fit observations of Tycho's supernova remnant~\citep{morlino+12}. Again using \verb|naima|, we then calculate the leptonic emission of a typical UFO from the weighted electron distributions, adding together the contribution of each shell to produce a cumulative SED at a given time step. As shown in Figure~\ref{fig:ufo_spectrum_full}, the resulting synchrotron emission always falls below the measured average radio emission of the galaxies and the inverse-Compton and bremsstrahlung emissions are a factor $>25$ fainter than the $\pi_0$ emission. Note that the inverse-Compton emission is estimated by assuming that electrons scatter off the cosmic microwave background and starlight approximated by a blackbody with temperature T = 3000 K. This emission remains a factor $>10$ below the $\pi_0$ one even with an artificially enhanced stellar radiation field of energy density 100 eV cm$^{-3}$. We also model the inverse-Compton emission assuming electrons scatter off the AGN photon field described in~\citet{Sazonov+04}, normalized to the bolometric luminosity of the AGN sample, and find that this emission remains a factor $>3$ below that produced by $\pi_0$-decay (and with a much softer spectrum above 30\,GeV).

Proton-proton interactions produce $\gamma$ rays with energy $E_{\gamma}\approx E_{\rm p}/10$~\citep{kelner2006}, and thus the observed $\gamma$-ray SED indicates a firm detection of CR protons with energies reaching at least as high as $\approx10^{12-13}$ eV. Within our hadronic emission model we derive that the maximum energy of protons accelerated at the forward shock is $\approx$10$^{17}$\,eV. This makes AGN winds a potential source of CRs with energies beyond the `knee' of the CR spectrum (i.e., $3\times10^{15}$\,eV) and also likely contributors to the IceCube neutrino flux~\citep{Aartsen:2013jdh}. 

\section{Discussion}
\label{sec:discussion}
This work has provided evidence for the existence of a new population of $\gamma$-ray emitters produced by AGN-driven outflows, which in the interaction with the ISM can create strong shocks able to energize charged particles potentially up to the transition region between Galactic and extragalactic CRs. These charged particles produce the observed $\gamma$ rays in the interaction with the ISM. According to our and other available models~\citep{Liu+18}, the observed emission can be explained by a forward shock that has traveled $\gtrsim$20\,pc from the central SMBH. As such, the $\gamma$-ray emission from UFOs may signal the onset of the wind-host interaction. We caution the reader, however, that our model relies on a one dimensional description of a galaxy and that it does not include the complex environment in the immediate vicinity of the SMBH. Nevertheless, our results are found to be in reasonably good agreement with previous predictions~\citep{lamastra2016galactic,Liu+18}. 
 
Most of the outflow energy is deposited in the bubble of hot gas rather than CRs. More precisely, this energy is a factor of $\sim$10 larger than what is transferred to accelerated CRs, which is reported in the right panel of Figure~\ref{fig:ufo_spectrum_full}. For an AGN with $\log L_{\rm Bol}$ (erg s$^{-1}$)=44.4 the timescale to transfer 10$^{56}$\,erg of energy to the bubble is about 3\,million years. This timescale would grow slightly for less powerful AGN. Sgr A*, the SMBH in the center of the Milky Way, has an estimated mass of 4 million M$_{\odot}$~\citep{gravity2019}, and was very likely in an active state up to a few hundred years ago~\citep{sunyaev1993,koyama96}. Adopting a bolometric luminosity of $\log L_{\rm Bol}$ (erg s$^{-1}$)=42.7 (about 1\,\% of its Eddington luminosity), we find that the timescale to deposit an energy of 10$^{56}$\,erg to the thermal gas is $\approx$10 million years. At that point in time, its $\gamma$-ray luminosity would be $\log L_{\rm \gamma}$ (erg s$^{-1}$)$\approx$38  and would decline modestly in a few hundred years after the end of the AGN activity \citep[see also][]{Liu+18}. This is in reasonable agreement with the  luminosity of the  bubbles discovered by {\it Fermi} in our galaxy ($\log L_{\rm \gamma}$ = 37.6 erg s$^{-1}$~\citep{su2010}) and the energetics of the thermal gas contained in the larger  bubbles ($\sim$10$^{56}$\,erg) recently discovered by eROSITA~\citep{predehl2020}. Thus, the {\it Fermi} and eROSITA bubbles may be the remnant of past UFO-like activity from the SMBH in the center of our galaxy.

It is important to note that the physical implications that can be inferred from the $\gamma$-ray detection are limited by the incompleteness of the current sample of UFOs, as well as the inherent uncertainty relating to the time variability of the UFOs. Indeed, detection of UFOs is limited to $\lesssim$50 AGN \citep{tombesi2010discovery,2016MNRAS.462..511K, 2020MNRAS.493.1088I}, which is by far not a complete sample.  Moreover, UFOs have been found in these AGN to vary with time and energy and this has been interpreted as a series of expanding shells \citep[see e.g.][]{king2015} rather than a continuous outflow like we have assumed here. Although we consider it unlikely, the variability may also impact the selection of a control sample as those AGN may not show a UFO precisely at the time when they were observed, but have otherwise an active UFO.
A solution to these issues will be provided with the more sensitive observations that the X-ray Imaging and Spectroscopy Mission \citep[XRISM,][]{xrism2020} and the Advanced Telescope for High-Energy Astrophysics (Athena\footnote{https://www.the-athena-x-ray-observatory.eu}) will provide.

\section{Summary and Conclusion}
\label{sec:summary_and_conclusion}
To search for the collective UFO emission, a stacking technique which has been used with success in the past~\citep{Ajello:2018sxm,paliya2019fermi,Ajello:2020zna} is adopted. Our sample consists of all radio-quiet UFOs with $z < 0.1$ and $v>0.1c$, which gives 11 sources in total. We model the UFO ($\gamma$-ray) spectrum with a power law, and we assume that the population can be characterized by an average flux and photon index. A fit to all the regions then optimizes these parameters. We find a TS of 30.1, which corresponds to a detection significance for the UFO emission of 5.1\,$\sigma$ (2 d.o.f.). The best-fit parameters are measured to be $\Gamma=-2.1\pm0.3$ and flux (1-800\,GeV) = $2.5^{+1.5}_{-0.9} \times 10^{-11} \ \mathrm{ph \ cm^{-2} \ {s^{-1}}}$. 

We performed several tests to confirm that the $\gamma$-ray emission is truly related to the presence of UFOs in this sample of galaxies. We employed a control sample of AGN with similar properties to those of the 11 galaxies used above, but lacking UFOs. This sample yields no detectable $\gamma$-ray emission with a (1-800\,GeV) flux upper limit of 8.8$\times10^{-12}$\, ph cm$^{-2}$ s$^{-1}$. We also use a sample of UFOs selected in a different  way~\citep{2020MNRAS.493.1088I} than our benchmark sample. These galaxies show a $\gamma$-ray signal whose parameters are in good agreement with those reported above. Moreover, adopting a control sample matched in X-ray flux, IR luminosity, radio flux and redshifts we can exclude that the observed $\gamma$-ray emission arises from star-formation activity or the presence of a weak jet. These tests allow us to conclude that the observed emission is associated to the presence of UFOs in these galaxies.

Observations of AGN winds have shown that AGN transfer a small fraction ($\sim$1$-$5\,\%) of their  bolometric luminosity to the winds. As our analysis indicates, a portion of this transferred luminosity in turn accelerates CRs and produces $\gamma$ rays. We find that AGN convert $\approx3\times10^{-4}$ of their bolometric luminosity into $\gamma$ rays. We also find that $\approx 4\times10^{-4}$ of the wind mechanical power is transferred to $\gamma$ rays. For comparison, in the Milky Way galaxy, supernova explosions transfer $\approx2\times 10^{-4}$ of their mechanical energy to $\gamma$ rays. This shows that AGN winds, if sustained for a few million years, can energize a large fraction of the CR population within a galaxy.

The physical model for the UFO SED is calculated by assuming that the $\gamma$-ray emission is dominated by hadronic processes resulting from diffusive shock acceleration. For typical UFO shock velocities and densities, a leptonic origin of the $\gamma$-ray emission is disfavored, in that inverse-Compton scattering and bremsstrahlung of relativistic electrons would produce  steeper $\gamma$-ray spectra with a  lower normalization. The observed $\gamma$-ray SED indicates a firm detection of CR protons with energies reaching at least as high as $\approx10^{12-13}$ eV.

Within our hadronic emission model we derive that on average the forward shock has traveled $\sim20-300$\ \ pc ($\sim65-980$ light years) away from the SMBH and that the maximum energy of protons accelerated at the forward shock is $\approx$10$^{17}$\,eV. This makes AGN winds a potential source of CRs with energies beyond the `knee' of the CR spectrum (i.e., $3\times10^{15}$\,eV) and also likely contributors to the IceCube neutrino flux~\citep{Aartsen:2013jdh,2018MNRAS.477.3469P}. Lastly, our results support the hypothesis that the \textit{Fermi} and eROSITA bubbles may be the remnant of past UFO-like activity from the SMBH in the center of our galaxy. 

 \section*{Acknowledgments}

M.~Ajello and C.~Karwin acknowledge support from NSF and NASA through grants AST-1715256 and 80NSSC18K1718. R.~Diesing and D.~Caprioli acknowledge the Eugene \& Niesje Parker Fellowship Fund, NASA (grants NNX17AG30G, 80NSSC18K1218, and 80NSSC18K1726) and the NSF (grants AST-1714658, AST-1909778). G.~Chartas acknowledges financial support from NASA grants 80NSSC20K0438 and 80NSSC19K095.

The \textit{Fermi} LAT Collaboration acknowledges generous ongoing support
from a number of agencies and institutes that have supported both the
development and the operation of the LAT as well as scientific data analysis.
These include the National Aeronautics and Space Administration and the
Department of Energy in the United States, the Commissariat \`a l'Energie Atomique
and the Centre National de la Recherche Scientifique / Institut National de Physique
Nucl\'eaire et de Physique des Particules in France, the Agenzia Spaziale Italiana
and the Istituto Nazionale di Fisica Nucleare in Italy, the Ministry of Education,
Culture, Sports, Science and Technology (MEXT), High Energy Accelerator Research
Organization (KEK) and Japan Aerospace Exploration Agency (JAXA) in Japan, and
the K.~A.~Wallenberg Foundation, the Swedish Research Council and the
Swedish National Space Board in Sweden.
Additional support for science analysis during the operations phase is gratefully
acknowledged from the Istituto Nazionale di Astrofisica in Italy and the Centre
National d'\'Etudes Spatiales in France. This work performed in part under DOE
Contract DE-AC02-76SF00515. Work at NRL is supported by NASA.

\bibliography{citations_master}{}
\bibliographystyle{aasjournal}

\end{document}